\documentclass[10pt]{emulateapj}
\usepackage{amsmath}
\usepackage{bm}
\usepackage{pslatex}
\usepackage{graphicx}
\usepackage{natbib}
\begin{document}

\title{Incorporating Kinetic Physics into a Two-Fluid Solar-Wind Model with Temperature Anisotropy and Low-Frequency Alfv\'en-Wave Turbulence}


\author{Benjamin D. G. Chandran\altaffilmark{1}, Timothy
  J. Dennis\altaffilmark{1}, Eliot Quataert\altaffilmark{2}, 
\& Stuart D. Bale\altaffilmark{3}}

\altaffiltext{1}{Space Science Center and Department of Physics,  University of New Hampshire, Durham, NH 03824;  benjamin.chandran@unh.edu, tim.dennis@unh.edu}

\altaffiltext{2}{Astronomy Department \& Theoretical Astrophysics
  Center, 601 Campbell Hall, The University of California, Berkeley,
  CA 94720; eliot@astro.berkeley.edu} 

\altaffiltext{3}{Physics Department \& Space Sciences Laboratory, 311
  Old LeConte Hall, The University of California, Berkeley,
  CA 94720; bale@ssl.berkeley.edu}

\begin{abstract}
  We develop a 1D solar-wind model that includes separate energy
  equations for the electrons and protons, proton temperature
  anisotropy, collisional and collisionless heat flux, and an
  analytical treatment of low-frequency, reflection-driven,
  Alfv\'en-wave (AW) turbulence. To partition the turbulent heating
  between electron heating, parallel proton heating, and perpendicular
  proton heating, we employ results from the theories of linear wave
  damping and nonlinear stochastic heating.  We account for mirror and
  oblique firehose instabilities by increasing the proton pitch-angle
  scattering rate when the proton temperature anisotropy exceeds the
  threshold for either instability.  We numerically integrate the
  equations of the model forward in time until a steady state is
  reached, focusing on two fast-solar-wind-like solutions. These
  solutions are consistent with a number of observations, supporting
  the idea that AW turbulence plays an important role in the origin of
  the solar wind.
\end{abstract} \keywords{solar wind --- Sun: corona --- turbulence --- waves}

\maketitle

\vspace{0.2cm} 
\section{Introduction}
\label{sec:intro}
\vspace{0.2cm} 

The first theory for the origin of the solar wind was developed by
E. N. Parker in the 1950s and 1960s. \cite{parker58,parker65} based
his analysis on a single-fluid, hydrodynamic model with electron
thermal conduction. Although he obtained smooth, transonic solutions
in agreement with some solar-wind observations, his work was unable to
account for the large flow velocities and proton temperatures measured
in fast-solar-wind streams near Earth. Subsequent studies generalized
the Parker model to include separate energy equations for the protons
and electrons~\citep{hartle68}, temperature
anisotropy~\citep{leer72,whang72}, super-radial expansion of the
magnetic field~\citep{holzer80}, collisionless heat
flux~\citep{hollweg74a,hollweg76}, and energy and momentum deposition by
Alfv\'en waves (AWs) and AW
turbulence~\citep{alazraki71,belcher71,hollweg73b,hollweg73a,tu87,tu88}.

The idea that AW turbulence plays an important role in the origin of
the solar wind was originally proposed by \cite{coleman68} and is
consistent with a number of observations. For example, data obtained
from the Solar Optical Telescope on the Hinode spacecraft reveal
the presence of ubiquitous, AW-like motions in the low corona carrying
an energy flux sufficient to power the solar
wind~\citep{depontieu07}. Pervasive AW-like fluctuations are also seen
at higher altitudes in the corona in observations from the Coronal
Multichannel Polarimeter of the National Solar
Observatory~\citep{tomczyk07}. Voyager, Helios, Wind, and other
spacecraft have measured broad-spectrum fluctuations in the magnetic
field, electric field, and flow velocity in the solar wind,
demonstrating that AW turbulence is present throughout the
interplanetary
medium~\citep{tumarsch95,goldstein95a,bale05}.
Moreover, the amplitudes of these turbulent fluctuations are
sufficient to explain the heating rates that have been inferred from
the proton and electron temperature
profiles~\citep{smith01,macbride08,cranmer09,stawarz09}.
Radio-scintillation observations of solar-wind density fluctuations
place upper limits on the turbulent heating rate in the solar wind
that are consistent with solar-wind heating by AW turbulence~(Harmon
\& Coles~2005; Chandran et al~2009; but see also
Spangler~2002). \nocite{harmon05,chandran09d,spangler02} In addition,
Faraday rotation of radio transmissions from the Helios spacecraft
show that magnetic-field fluctuations in the corona at heliocentric
distances between $2R_{\sun}$ and $15R_{\sun}$ are consistent with
models in which the solar wind is driven by AW
turbulence~\citep{hollweg82b,hollweg10}.

However, it is not clear that heating by AW turbulence can explain
observations of ion temperature anisotropies.  Measurements from the
Helios spacecraft have shown that $T_{\perp \rm p} > T_{\parallel \rm
  p}$ in the core of the proton velocity distribution in
low-$\beta_{\parallel \rm p}$ fast-wind streams, where
\begin{equation}
\beta_{\parallel \rm p} = \frac{8 \pi n k_{\rm B} T_{\parallel \rm p}}{B^2},
\label{eq:betapar} 
\end{equation} 
$n$ is the proton density, $k_{\rm B}T_{\parallel \rm p}/2$ is the
average energy per proton in thermal motions parallel to the magnetic
field~$\bm{B}$, and $k_{\rm B} T_{\perp \rm p}$ is the average energy
per proton in thermal motions perpendicular
to~$\bm{B}$~\citep{marsch04}.  Similarly, remote observations from the
Ultraviolet Coronagraph Spectrometer (UVCS) on the Solar
and Heliospheric Observatory have shown that $T_\perp \gg
T_\parallel$ for ${\rm O}^{+5}$ ions in coronal
holes~\citep{kohl98,li98,antonucci00}.  These observations
pose a challenge to solar-wind models based on AW turbulence because
the AW energy cascade is anisotropic, transporting AW energy primarily
to small scales measured perpendicular to~$\bm{B}$ and only weakly to
small scales measured parallel to~$\bm{B}$~\citep{shebalin83}. As a
consequence, very little energy cascades to high frequencies
comparable to the ion cyclotron frequencies at which waves can dissipate
via resonant cyclotron interactions~\citep{quataert98}, which are the
only route to perpendicular ion heating in a collisionless plasma
within the framework of quasilinear theory~\citep{stix92}.

A number of studies have gone beyond quasilinear theory to show that
low-frequency AW turbulence can lead to perpendicular ion heating even
in the absence of a cyclotron
resonance~\citep{mcchesney87,chen01,johnson01,dmitruk04,voitenko04,bourouaine08,parashar09,chandran10a,chandran10b}.
In this paper, we incorporate results from one of these
studies~\citep{chandran10a} into a quantitative solar-wind model to
investigate the extent to which low-frequency AW turbulence can
explain the observations of anisotropic proton temperatures discussed
above. The model we have developed also includes the non-WKB
reflection of AWs, proton and electron heat flux in both the
collisional and collisionless regimes, and enhanced pitch-angle
scattering when the proton temperature anisotropy is sufficiently
large to excite mirror, cyclotron, or firehose instabilities.  We
describe this model in detail in Section~\ref{sec:fluxtube} and
discuss the numerical method we use to solve the equations of the
model in Section~\ref{sec:method}.  In Section~\ref{sec:solution} we
present and analyze a steady-state, fast-wind-like solution and
compare this solution to a number of observations. In
Section~\ref{sec:discussion} we present a second steady-state solution
and discuss our results. We conclude in Section~\ref{sec:conclusion}
and summarize the derivation of the equations of our model in the
appendix.

\vspace{0.2cm} 
\section{Mathematical Framework}
\label{sec:fluxtube} 
\vspace{0.2cm} 

In the solar wind, the Debye length~$\lambda_{\rm D}$ and proton
gyroradius~$\rho_{\rm p}$ are vastly smaller than the length scales
over which the bulk solar-wind properties vary appreciably. Because of
this, the large-scale structure of solar-wind plasma can be rigorously
described by taking the limit of the Vlasov and Maxwell equations in
which $\rho_{\rm p} \rightarrow 0$ and $\lambda_{\rm D} \rightarrow
0$. This limit, described in more detail in the appendix, is
some times referred to as Kulsrud's collisionless magnetohydrodynamics
\citep{kulsrud83}. This name is somewhat misleading, in that the
resulting description is still kinetic in nature. In particular, two
of the variables in Kulsrud's theory are the reduced proton and
electron distribution functions, $f_{\rm p}$ and $f_{\rm e}$, which
are independent of the gyrophase angle in velocity space and evolve in
time according to the guiding-center Vlasov equation
(Equation~(\ref{eq:dfdt})).

We take Kulsrud's collisionless magnetohydrodynamics (MHD) as the
starting point for our study, including the simplifying assumption
that the plasma consists of just protons and electrons.  We then
follow \cite{snyder97} in adding a collision operator to the
guiding-center Vlasov equation and taking velocity moments of this
equation to obtain a hierarchy of fluid equations, as described in
more detail in the appendix. To close this set of fluid equations for
the protons, we set $f_{\rm p}$ equal to a bi-Maxwellian when
evaluating the various fourth moments of~$f_{\rm p}$. This approach
offers a simple and, we expect, reasonably accurate way to solve for
the proton heat flux in the presence of temperature anisotropy in both
collisional and collisionless conditions. The same general 
equations (with differing treatments of the collision terms) were
derived in different ways by  \cite{endeve01},  \cite{liesvendsen01}, and~\cite{ramos03}.
For the electrons, we adopt the simplifying assumption
that $f_{\rm e}$ is isotropic in velocity space and close the fluid
equations by specifying the electron heat flux in terms of lower-order
moments of~$f_{\rm e}$.  The most general versions of the resulting
equations are given in the appendix. In this section, we specialize
these equations to our 1D solar-wind model and add terms to
incorporate heating by AW wave turbulence, acceleration by the AW
pressure force, and temperature isotropization by firehose and mirror
instabilities.

To simplify the analysis, we neglect the Sun's rotation, take the
background magnetic field to be fixed, and solve the fluid equations
within a narrow, open magnetic flux tube centered on a radial magnetic
field line. We follow \cite{kopp76} in taking the cross sectional area
of this flux tube to be
\begin{equation}
a = a_{\sun} \left(\frac{r}{R_{\sun}}\right)^2 f,
\label{eq:defa} 
\end{equation}
where $r$ is heliocentric distance,
\begin{equation}
f = \frac{f_{\rm max} e^{(r-R_1)/\sigma} + f_1}{e^{(r-R_1)/\sigma} + 1},
\label{eq:deff} 
\end{equation} 
$f_1 = 1 - (f_{\rm max} - 1) \exp[(R_{\sun} - R_1)/\sigma]$, and
$a_{\sun}$, $f_{\rm max}$, $R_1$, and~$\sigma$ are constants.  The
function~$f$ increases from~1 at $r=R_{\sun}$ to $f_{\rm max}$ at
$r\gg R_1$, with most of the variation in $f$ occurring between $r =
R_1 - \sigma$ and $r= R_1 + \sigma$.  To reduce the number of free
parameters, we set
\begin{equation}
\sigma = R_1.
\label{eq:sigmaR1} 
\end{equation} 
The inner radius of our model
corresponds to the coronal base just above the transition region. We
set this radius equal to~$R_{\sun}$, neglecting the thickness of the
chromosphere and transition region. By flux conservation, the strength
of the magnetic field $\bm{B}$ satisfies
\begin{equation}
B = \frac{ B_{\sun} \, a_{\sun}}{a},
\label{eq:B}
\end{equation} 
where $B_{\sun}$ is the magnetic field strength at the coronal base. We
take the cross sectional area at the coronal base, $a_{\sun}$, to be $ \ll
R_\sun^2$ and $\ll R_1^2$, so that the flux tube is thin. As a consequence, the
magnetic field direction is approximately radial everywhere within the
flux tube. We thus set
\begin{equation}
\bm{\hat{b}} \cdot \bm{\nabla} \rightarrow  \frac{\partial}{\partial r},
\label{eq:bdotgrad} 
\end{equation} 
where we have taken the magnetic field to be pointing away from the Sun. 
The condition $\bm{\nabla } \cdot \bm{B} = 0$ then gives
\begin{equation}
\bm{\nabla } \cdot \bm{\hat{b}} = \frac{1}{a} \frac{\partial a}{\partial r}.
\label{eq:divbhat} 
\end{equation} 
We take the solar-wind outflow velocity to be everywhere parallel
to the magnetic field,
\begin{equation}
\bm{U} = U \bm{\hat{b}},
\label{eq:upar} 
\end{equation} 
and define the Lagrangian time derivative
\begin{equation}
\frac{d}{dt} = \frac{\partial }{\partial t} + U \frac{\partial
}{\partial r}.
\label{eq:lagder1} 
\end{equation} 

There are eight dependent variables in our model: the proton (or
electron) number density~$n$, the proton (or electron) outflow
velocity~$U$, the electron temperature~$T_{\rm e}$, the perpendicular
and parallel proton temperatures $T_{\perp \rm p}$ and $T_{\parallel
  \rm p}$, the proton heat fluxes $q_{\perp \rm p}$ and $q_{\parallel
  \rm p}$, and the energy density of AWs propagating away
from the Sun~${\cal E}_{\rm w}$. The
proton heat flux~$q_{\perp \rm p}$ is a flow along~$\bm{B}$ of
perpendicular proton kinetic energy; no heat flows across the magnetic
field in the model. The eight dependent variables of the model depend upon time~$t$
and a single spatial coordinate~$r$ and satisfy
the following eight equations:
\begin{equation} 
\frac{d n}{d t}  = -\, \frac{n}{a} \frac{\partial}{\partial r} \left( a
  U\right),
 \label{eq:cont1} 
\end{equation}
\vspace{0.2cm} 
\[ \frac{d U}{dt} = \,-\, \frac{k_{\rm B}}{\rho} \frac{\partial }{\partial r}\left[ n(T_{\rm e} + T_{\parallel \rm p})\right]  + \frac{k_{\rm B}(T_{\perp \rm p}
- T_{\parallel \rm p})}{m_{\rm p} a} \frac{\partial a}{\partial r}
\] 
\begin{equation}
- \, \frac{G M_{\sun}}{r^2} - \frac{1}{2\rho} \frac{\partial {\cal E}_{\rm w}}{\partial r},
\label{eq:momentum}
\end{equation}
\vspace{0.2cm} 
\begin{equation} 
\frac{3}{2}n^{5/3} k_{\rm B} \frac{d}{dt}\left(\frac{T_{\rm e}}{n^{2/3}}\right) =
Q_{\rm
    e} - \frac{1}{a} \frac{\partial}{\partial r} (a q_{\rm e}) + 3\nu_{\rm pe} n k_{\rm B} (T_{\rm p} - T_{\rm e})
\label{eq:dTedt},
\end{equation}
\vspace{0.2cm} 
\[
B n k_{\rm B} \frac{d}{dt}\left( \frac{T_{\perp \rm p}}{B}\right) =
Q_{\perp \rm p} - \frac{1}{a^2}\frac{\partial }{\partial r}\left(a^2
  q_{\perp \rm p}\right)+ \frac{1}{3} \nu_{\rm p} n k_{\rm B}(T_{\parallel \rm p} - T_{\perp \rm p})
\]
\begin{equation}
+ \,2\nu_{\rm pe} n k_{\rm B} (T_{\rm e} - T_{\perp \rm p}),
\label{eq:dTperpdt} 
\end{equation} 
\vspace{0.2cm} 
\[
\frac{n^3 k_{\rm B}}{2B^2} \frac{d}{dt}\left(\frac{B^2
    T_{\parallel \rm p}}{n^2}\right) =  Q_{\parallel \rm p} -
\frac{1}{a} \frac{\partial }{\partial r} ( a q_{\parallel \rm p})
+ \frac{q_{\perp \rm p}}{a}\,\frac{\partial a}{\partial r}
\]
\begin{equation}
+\, \frac{1}{3} \nu_{\rm p} n k_{\rm B}(T_{\perp \rm p} - T_{\parallel\rm p}) 
+ \nu_{\rm pe} n k_{\rm B}(T_{\rm e} - T_{\parallel \rm  p}),
\label{eq:dTpardt}
\end{equation} 
\vspace{0.2cm}
\[
n^2 \frac{d}{dt}\left(\frac{q_{\perp \rm p}}{n^2}\right) = - \, \frac{n k_{\rm B}^2 T_{\parallel
    \rm p}}{m_{\rm p}} \frac{\partial T_{\perp \rm p}}{\partial r}
+
\frac{n k_{\rm B}^2 T_{\perp \rm p}(T_{\perp \rm p} - T_{\parallel \rm
    p})}{m_{\rm p} a} \frac{\partial a}{\partial r}
\]
\begin{equation}
 - \,
\nu_{\rm p} q_{\perp \rm p},
\label{eq:dqperpdt} 
\end{equation} 
\vspace{0.2cm} 
\begin{equation}
\frac{n^4}{B^3} \frac{d}{dt}\left(\frac{B^3 q_{\parallel \rm p}}{n^4}\right) = - \frac{3
  n k_{\rm B}^2 T_{\parallel \rm p}}{2m_{\rm p}} \frac{\partial
  T_{\parallel \rm p}}{\partial r} -  \nu_{\rm p} q_{\parallel \rm p},
\label{eq:dqpardt} 
\end{equation} 
\vspace{0.2cm} 
\noindent and \citep{dewar70}
\begin{equation} 
\frac{\partial {\cal E}_{\rm w} }{\partial t} + \frac{1}{a} \frac{\partial }{\partial r}\left[a (U + v_{\rm A}){\cal E}_{\rm w}\right] + \frac{{\cal E}_{\rm w}}{2a}\frac{\partial}{\partial r}(a U) = - Q
\label{eq:dEwdt} ,
\end{equation} 
where
\begin{equation}
v_{\rm A}  = \frac{B}{\sqrt{4 \pi \rho}}
\label{eq:vA} 
\end{equation} 
is the Alfv\'en speed,
$\rho$ is the mass density,  and $M_{\sun}$ is the mass of the Sun.
Since we are treating the solar wind as a proton-electron
plasma, we ignore the contribution of alpha particles and other
particle species to~$\rho$ and set
\begin{equation}
\rho = m_{\rm p} n.
\label{eq:rho0}
\end{equation} 
The quantities $Q_{\rm e}$, $Q_{\perp
  \rm p}$, and $Q_{\parallel \rm p}$ are, respectively, the electron
heating rate, the perpendicular proton heating rate, and the parallel
proton heating rate per unit volume
from the dissipation of AW turbulence (Section~\ref{sec:stoch}), 
\begin{equation}
Q = Q_{\rm e} + Q_{\perp \rm p} + Q_{\parallel \rm p}
\label{eq:defQ} 
\end{equation} 
is the total turbulent heating rate per unit volume, and
$q_{\rm e}$ is the electron heat flux (Section~\ref{sec:eheatflux}).  
The quantity
\begin{equation}
\nu_{\rm pe} = \frac{ 4 \sqrt{2 \pi m_{\rm e}} \,e^4 n \ln \Lambda }{3 m_{\rm p} (k_{\rm B}T_{\rm e})^{3/2}}
\label{eq:nupe} 
\end{equation} 
is the Coulomb collision frequency for energy exchange between protons
and electrons~\citep{schunk75}, where $m_{\rm p}$ and $m_{\rm e}$ are
the proton and electron masses
and $\ln \Lambda$ is the Coulomb logarithm, which we take to be~23.
We set 
\begin{equation}
\nu_{\rm p} = \nu_{\rm pp, C} + \nu_{\rm inst},
\label{eq:nupp} 
\end{equation} 
where
\begin{equation}
\nu_{\rm pp, C} = \frac{ 4 \sqrt{\pi} \, e^4 n \ln \Lambda }{3 \sqrt{m_{\rm p}} (k_{\rm B}T_{\rm p})^{3/2}}
\label{eq:nuppC} 
\end{equation} 
is the proton-proton Coulomb collision frequency~\citep{schunk75}, 
$\nu_{\rm inst}$ is a scattering rate associated with small-scale plasma waves that are excited when the proton temperature anisotropy becomes sufficiently large (Section~\ref{sec:inst}), and
\begin{equation}
T_{\rm p} = \frac{2 T_{\perp \rm p} + T_{\parallel \rm p}}{3}.
\label{eq:Tp} 
\end{equation} 

Equation~(\ref{eq:cont1}) expresses the conservation of mass in our 1D
model. Equation~(\ref{eq:momentum}) is the same as the momentum
equation in Kulsrud's collisionless MHD
(Equation~(\ref{eq:momentum0})), except that we have added the
gravitational acceleration and the AW pressure force
\citep{dewar70}.  In the absence of turbulent heating, heat flow, and
collisions, the right-hand sides of Equations~(\ref{eq:dTedt}) through
(\ref{eq:dTpardt}) vanish. In this case, the Lagrangian time
derivative of the electron specific entropy ($\propto \ln (T_{\rm
  e}/n^{2/3}))$ vanishes, and the protons obey the double-adiabatic
theory of Chew, Goldberger, \& Low~(1956). \nocite{chew56} When
$\nu_{\rm p}$ becomes sufficiently large, $T_{\perp \rm p} \simeq
T_{\parallel \rm p}$ and the collisional terms in
Equations~(\ref{eq:dqperpdt}) and (\ref{eq:dqpardt}) (which are
$\propto \nu_{\rm p}$) are much larger than the left-hand sides of
these equations. In this limit, $q_{\perp \rm p}$ and $q_{\parallel
  \rm p}$ are determined from Equations~(\ref{eq:dqperpdt}) and
(\ref{eq:dqpardt}) by balancing the collisional terms against the
source terms (which contain $T_{\perp \rm p}$ and/or $T_{\parallel \rm
  p}$ but not $q_{\perp \rm p}$ or $q_{\parallel \rm p}$), and the
total proton heat flux $q_{\perp \rm p} + q_{\parallel \rm p}$
becomes approximately equal to the proton heat flux in collisional
transport theory~\citep{bra65}.

Upon multiplying Equation~(\ref{eq:momentum}) by~$\rho U$ and adding
the resulting equation to the sum of Equations~(\ref{eq:dTedt}),
(\ref{eq:dTperpdt}), (\ref{eq:dTpardt}), and~(\ref{eq:dEwdt}), we
obtain a total energy equation,
\begin{equation}
\frac{\partial {\cal  E}_{\rm tot} }{\partial t} + \frac{1}{a} \frac{\partial }{\partial r}(a F_{\rm tot}) = 0,
\label{eq:total_energy_eqn} 
\end{equation} 
where 
\begin{equation}
{\cal E}_{\rm tot} = \frac{\rho U^2}{2} - \frac{G M_{\sun}\rho}{r} + n k_{\rm B} \left(
\frac{3T_{\rm e}}{2} + T_{\perp \rm p} + \frac{T_{\parallel \rm p}}{2}
\right) + {\cal E}_{\rm w}
\label{eq:Etot} 
\end{equation} 
is the total energy density, and
\[
F_{\rm tot} = \frac{ \rho U^3}{2}  - \frac{U G M_{\sun}\rho }{r}
+  U n k_{\rm B} \left( \frac{5 T_{\rm e}}{2} + T_{\perp \rm p} + \frac{ 3T_{\parallel \rm p}}{2}\right)
\]
\begin{equation}
+ \, q_{\rm e} + q_{\perp \rm p} + q_{\parallel \rm p}
+ \left( \frac{3 U }{2} + v_{\rm A} \right){\cal E}_{\rm w}
\label{eq:Gammatot} 
\end{equation} 
is the total energy flux. In steady state, $aF_{\rm tot}$ is
independent of~$r$, and the total flow of energy into the base of the
flux tube equals the total flow of energy through the
flux-tube cross section at all other radii.

\subsection{Electron Heat Flux}
\label{sec:eheatflux} 
Close to the Sun, $n$ is sufficiently large that the electron Coulomb mean free path,
\begin{equation}
\lambda_{\rm mfp} = \frac{\sqrt{k_{\rm B} T_{\rm e}/m_{\rm e}}}{\nu_{\rm e}},
\label{eq:lambda} 
\end{equation} 
is much shorter than the radial distance $l_{T} = T_{\rm
  e}/(\partial T_{\rm e}/\partial r)$ over which $T_{\rm e}$ varies
appreciably. The quantity
\begin{equation}
\nu_{\rm e} = 2.9 \times 10^{-6} \left(\frac{n}{1 \mbox{ cm}^{-3}}\right) \left(\frac{k_{\rm B}T_{\rm e}}{1 \mbox{ eV}}\right)^{-3/2} \ln \Lambda \mbox{ s}^{-1}
\label{eq:nuee} 
\end{equation} 
is the electron collision frequency~\citep{plasmaformulary83}.
We expect the electron heat flux in this near-Sun region to be
approximately equal to the Spitzer value \citep{spitzer53},
\begin{equation}
\bm{q}_{\rm e, S} = - \kappa_{\rm e0} T_{\rm e}^{5/2} (\bm{\hat{b}} \cdot \bm{\nabla} T_{\rm e}) \bm{\hat{b}},
\label{eq:qSp} 
\end{equation} 
where 
\begin{equation}
\kappa_{\rm e0} = \frac{1.84 \times 10^{-5} }{\ln \Lambda} \mbox{ erg} \mbox{ s}^{-1} \mbox{ K}^{-7/2} \mbox{ cm}^{-1}.
\label{eq:kappe0} 
\end{equation} 
Farther from the Sun, $\lambda_{\rm mfp}\gtrsim l_T$, and $q_{\rm e}$
deviates from the Spitzer value. We follow \cite{hollweg74a,hollweg76}
in taking the collisionless heat flux in this region to be
approximately
\begin{equation}
\bm{q}_{\rm e, H} =  \frac{3}{2} \alpha_{\rm H} U  n k_{\rm B} T_{\rm e} \bm{\hat{b}} ,
\label{eq:qHollweg} 
\end{equation} 
where $\alpha_{\rm H}$ is a constant that we treat as a free
parameter.  \cite{hollweg76} argued that the transition between the
collisional and collisionless regimes occurs at the radius at which
$\lambda_{\rm mfp} \simeq 0.5 r$. To interpolate smoothly between the
two regimes, we set the electron heat flux equal to
\begin{equation}
\bm{q}_{\rm e} = \psi \bm{q}_{\rm e, H} + (1-\psi) \bm{q}_{\rm e, S},
\label{eq:qe} 
\end{equation} 
where
\begin{equation}
\psi = \frac{(r/r_{\rm H})^2}{1 + (r/r_{\rm H})^2},
\label{eq:defpsi} 
\end{equation} 
and $r_{\rm H}$ is a constant that we choose to coincide
with the radius at which $\lambda_{\rm mfp}  = 0.5 r$.
For the numerical solutions presented in Sections~\ref{sec:solution}
and~\ref{sec:swani_cyclotron}, we set
$r_{\rm H} = 5 R_{\sun}$, and confirm post facto that $\lambda
\simeq 0.5 r$ at $r=r_{\rm H}$ (see Figure~\ref{fig:swq}).

\subsection{Proton Pitch-Angle Scattering from Firehose and Mirror Instabilities}
\label{sec:inst} 

If the proton temperature-anisotropy ratio
\begin{equation}
R = \frac{T_{\perp \rm p}}{T_{\parallel \rm p}}
\label{eq:defR} 
\end{equation} 
becomes either too large or too small, the plasma becomes
unstable. Spacecraft measurements show that the values of~$R$ found in
the solar wind are bounded from below by the instability threshold of
the oblique firehose mode and from above by the instability threshold
of the mirror mode~\citep{kasper02,hellinger06,bale09}. In particular,
most of the measured values of $R$ correspond to plasma parameters for
which $\gamma_{\rm max} < 10^{-3}\Omega_{\rm p}$, where~$\gamma_{\rm
  max}$ is the maximum growth rate of the oblique firehose or mirror
instability and~$\Omega_{\rm p}$ is the proton cyclotron frequency.
The value of~$R$ for which~$\gamma_{\rm max} = 10^{-3} \Omega_{\rm p}$
is approximately 
 \begin{equation}
 R_{\rm m} = 1 + 0.77(\beta_{\parallel \rm p} + 0.016)^{-0.76} 
\label{eq:Rmirror} 
\end{equation} 
for the mirror instability, 
and approximately
\begin{equation}
 R_{\rm f} = 1 - 1.4(\beta_{\parallel \rm p} + 0.11)^{-1} 
\label{eq:Rfirehose} 
\end{equation} 
for the oblique firehose instability \citep{hellinger06}.

Presumably, when the plasma becomes unstable, small-scale
electromagnetic fluctuations grow and enhance the proton pitch-angle
scattering rate, preventing the temperature anisotropy from increasing
further. We incorporate this effect into our model through the
term~$\nu_{\rm inst}$ in Equation~(\ref{eq:nupp}), with
\begin{equation}
\nu_{\rm inst} = \nu_0 \exp\left[ \frac{12(R - R_{\rm m})}{R_{\rm m}}\right]
+ \nu_0 \exp\left[ \frac{12(\overline{ R}_{\rm f} -
      R)}{ \overline{R}_{\rm f}}\right],
\label{eq:nuinst} 
\end{equation} 
$\nu_0 = 0.02\sqrt{G M_{\sun}/R_{\sun}^3}$, and $\overline{ R}_{\rm f} =
\max ( R_{\rm f}, 10^{-6})$.  A similar approach was employed by
\cite{sharma06a} in numerical simulations of accretion flows around
black holes.

\subsection{Alfv\'en Wave Turbulence}
\label{sec:Alfven} 

The Sun launches different types of waves that propagate outward into
the solar atmosphere. In our model, we retain only the non-compressive
Alfv\'en wave~(AW), in part for simplicity and in part because the AW
is the most promising wave type for transporting energy over large
distances into the corona and solar
wind~\citep{barnes66,velli89,matthaeus99b,suzuki05,cranmer05}. For AW
fluctuations, the fluctuating velocity vector $\delta \bm{v}$ and
magnetic field vector~$\delta \bm{B}$ lie in the plane perpendicular
to~$\bm{B}_0$. We define the Elsass\"er variables
\begin{equation}
\bm{z}^\pm = \delta \bm{v} \mp \frac{\delta \bm{B}}{\sqrt{4 \pi \rho}},
\end{equation} 
and, as mentioned previously, take~$\bm{B}_0$ to point away from the
Sun. In the small-amplitude limit, $\bm{z}^+$ fluctuations are AWs
that propagate with an outward radial velocity of $U + v_{\rm A}$,
while the $\bm{z}^-$ fluctuations are AWs that propagate with a radial
velocity $U-v_{\rm A}$. Near the Sun, $U<v_{\rm A}$ and~$z^-$
fluctuations propagate towards smaller~$r$. 

To a good approximation, AWs in the solar corona and solar wind can be
described within the framework of
reduced~MHD~\citep{kadomtsev74,strauss76,zank92,schekochihin09}.  In
reduced MHD, the outward-propagating $z^+$ waves generated by the Sun
do not interact with one another. However, we assume that most of the
AW energy is at periods of tens of minutes to hours, which makes the
wavelengths in the radial direction sufficiently long that the AWs
undergo significant non-WKB reflection, converting some of the $z^+$
waves into~$z^-$ waves~\citep{heinemann80, velli93}. Interactions
between $z^+$ and $z^-$ fluctuations then cause wave energy to cascade
from large scales to small scales
\citep{iroshnikov63,kraichnan65,velli89,
  matthaeus99b,vanballegooijen11}. At sufficiently small scales, the
$z^\pm$ fluctuations dissipate. Although some reflection occurs, we
assume that
\begin{equation} 
z^- \ll z^+
\label{eq:zpzm} 
\end{equation} 
and neglect the contribution of~$z^-$ to the wave energy
density~${\cal E}_{\rm w}$, which is then given by \begin{equation}
{\cal E}_{\rm w} = \frac{\rho (z^+ _{\rm rms})^2}{4}, 
\label{eq:Ew} 
\end{equation}
where $z^+_{\rm rms}$ is the rms amplitude of~$z^+$ fluctuations.

To describe the cascade of wave energy in the presence of wave
reflections, we adopt the phenomenological model of \cite{dmitruk02},
which was later extended by \cite{chandran09c} to account for the
solar-wind outflow velocity. The essence of these models is to balance
the rate at which $z^-$ waves are produced by wave reflections against
the rate at which the $z^-$ waves cascade and dissipate via
interactions with $z^+$ waves. This balance leads to the following estimate for
the rms amplitude of $z^-$~\citep{chandran09c}:
\begin{equation}
z^-_{\rm rms} = \frac{L_{\perp} (U + v_{\rm A})}{v_{\rm A}} \,\left| \frac{\partial v_{\rm A}}{\partial r}\right|,
\label{eq:zminus} 
\end{equation} 
where $L_{\perp}$ is the correlation length (outer scale) of the
Alfv\'enic fluctuations in the plane perpendicular to~$\bm{B}_0$.  The
rate at which energy cascades and dissipates per unit volume is then
\begin{equation} 
Q = \frac{c_{\rm d}\, \rho\, z_{\rm rms}^- \,(z_{\rm rms}^+)^2}{4 L_{\perp}},
\label{eq:Q} 
\end{equation} 
where $c_{\rm d}$ is a dimensionless number. Since our
estimate of~$z^-_{\rm rms}$ is proportional to~$L_{\perp}$, the value
of~$Q$ in equation~(\ref{eq:Q}) is independent of~$L_\perp$.  Because
of Equation~(\ref{eq:zpzm}), we have omitted a term $\propto z_{\rm
  rms}^+ \,(z_{\rm rms}^-)^2$ that is some times included in the
turbulent heating rate~\citep{hossain95}.

\subsection{Proton and Electron Heating Rates}
\label{sec:stoch} 

In Equation~(\ref{eq:dEwdt}), the rate~$Q$ at which energy is drained
from the AWs equals the energy cascade rate given in
Equation~(\ref{eq:Q}), which is determined by the ``large-scale
quantities'' $z^+_{\rm rms}$, $z^-_{\rm rms}$, and~$L_\perp$.  All of
the AW energy that cascades to small scales dissipates, contributing
to turbulent heating, but the way that~$Q$ is apportioned
between~$Q_{\rm e}$, $Q_{\perp \rm p}$, and~$Q_{\parallel \rm p}$
depends upon the mechanisms that dissipate the fluctuations at length
scales~$\ll L_{\perp}$. In this section, we describe how we divide the
turbulent heating power between ~$Q_{\rm e}$, $Q_{\perp \rm p}$,
and~$Q_{\parallel \rm p}$ using results from the theories of linear
wave damping and nonlinear stochastic heating.

Nonlinear interactions between counter-propagating AWs cause AW energy
to cascade primarily to larger~$k_\perp$ and only weakly to
larger~$|k_\parallel|$, where $k_\perp$ and $k_\parallel$ are
wavevector components perpendicular and parallel to~$\bm{B}_0$
\citep{shebalin83,goldreich95,ng96,galtier00}. This cascade does not
transfer AW energy efficiently to higher frequencies (the AW frequency
being~$k_\parallel v_{\rm A}$), and thus cyclotron damping is not an
important dissipation mechanism for the anisotropic AW
cascade~\citep{cranmer03,howes08a}. There may be other mechanisms in
the solar wind that generate AWs with sufficiently high frequencies
that the AWs undergo cyclotron damping~\citep{leamon98a,hamilton08},
such as a turbulent cascade involving compressive
waves~\citep{chandran05a,chandran08b,yoon08} or instabilities driven
by proton or alpha-particle
beams~\citep{gomberoff96,hellinger11}. However, we do not account for
these possibilities in our model.

When AW energy cascades to~$k_\perp \rho_{\rm p} \simeq 1$, the
cascade transitions to a kinetic Alfv\'en wave (KAW)
cascade~\citep{bale05,howes08a,howes08b,schekochihin09,sahraoui09},
and the KAW fluctuations undergo Landau damping and transit-time
damping~\citep{quataert98,gruzinov98,leamon99} and dissipation via
stochastic heating~\citep{mcchesney87,chen01,johnson01}. Some of the
turbulent energy dissipates at~$k_\perp \rho_{\rm p} \simeq 1$, and
some of the turbulent energy cascades to, and then dissipates at,
smaller scales. Before describing the details of how we incorporate
dissipation into our model, we first summarize our general approach.
We make the approximation that the dissipation occurs in two distinct wavenumber
ranges: $k_\perp \rho_{\rm p} \sim 1$ and $k_\perp \rho_{\rm p} \gg
1$. We divide the total dissipation power between these two wavenumber
ranges by comparing the energy cascade time scale and damping time
scale at $k_\perp \rho_{\rm p} = 1$ (see
Equation~(\ref{eq:Gamma}) below). We divide the power that is
dissipated at $k_\perp \rho_{\rm p} \sim 1$ between~$Q_{\rm e}$,
$Q_{\perp \rm p}$, and~$Q_{\parallel \rm p}$ by comparing the damping
rates at $k_\perp \rho_{\rm p} = 1$ associated with three
different dissipation mechanisms, each of which contributes primarily
to either~$Q_{\rm e}$, $Q_{\perp \rm p}$, or~$Q_{\parallel \rm p}$. We
then assume that all of the power that dissipates at~$k_\perp \gg
\rho_{\rm p}^{-1}$ does so via interactions with electrons, thereby
contributing to~$Q_{\rm e}$.

\begin{figure}[h]
\centerline{
\includegraphics[width=7cm]{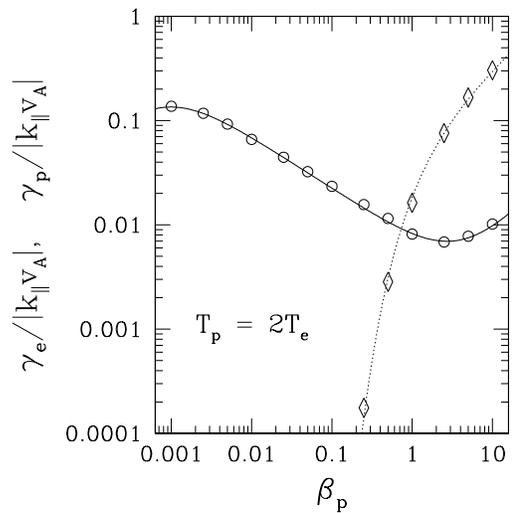}
}
\caption{Solid line gives the electron contribution to the KAW damping
  rate at~$k_\perp \rho_{\rm p} = 1$ from Equation~(\ref{eq:gammae}),
  and the dotted line gives the proton contribution to the KAW damping
  rate at~$k_\perp \rho_{\rm p} = 1$ from
  Equation~(\ref{eq:gammap}). The circles and diamonds are,
  respectively, the electron and proton
  contributions to the KAW damping rate at $k_\perp \rho_{\rm p} = 1$ 
in our numerical solutions to the full hot-plasma dispersion relation
for Maxwellian electrons and protons.
\label{fig:gamma_e_p} }\vspace{0.5cm}
\end{figure}

We define~$\gamma_{\rm e}$ and $\gamma_{\rm p}$ to be the electron and
proton contributions to the linear damping rate of KAWs at $k_\perp
\rho_{\rm p} = 1$, where $\rho_{\rm p}$ is the proton gyroradius.
Using a numerical code that solves the full hot-plasma dispersion
relation~\citep{quataert98}, we have calculated $\gamma_{\rm e}$ and
$\gamma_{\rm p}$ for a range of plasma parameters, assuming isotropic
proton and electron temperatures.  For $10^{-3} < \beta_{\rm p} < 10$,
$1 \lesssim T_{\rm p}/T_{\rm e} \lesssim 5$, and $|k_\parallel v_{\rm
  A} |\ll \Omega_{\rm p}$, our results are well approximated by the
following formulas:
\begin{equation}
\frac{  \gamma_{\rm e}}{|k_\parallel v_{\rm A}|} = 0.01\left(\frac{T_{\rm e}}{T_{\rm p} 
      \beta_{\rm p}}\right)^{1/2}
\left[  \frac{ 1 + 0.17 \beta_{\rm
      p}^{1.3} }{1 + (2800\beta_{\rm e})^{-1.25}}\right]
\label{eq:gammae} 
\end{equation} 
and
\begin{equation}
\frac{\gamma_{\rm p}}{|k_\parallel v_{\rm A}|} =  0.08  \left(\frac{T_{\rm e}}{T_{\rm
      p}}\right)^{1/4} \beta_{\rm p}^{0.7} \exp\left(- \,
  \frac{1.3}{\beta_{\rm p}}\right) ,
\label{eq:gammap} 
\end{equation} 
where $\beta_{\rm p} = 8\pi n k_{\rm B} T_{\rm p}/B_0^2$
and $\beta_{\rm e} = 8\pi n k_{\rm B} T_{\rm e}/B_0^2$.
In Figure~\ref{fig:gamma_e_p}, we compare 
Equations~(\ref{eq:gammae}) and (\ref{eq:gammap}) 
with our numerical solutions for the case in which
$T_{\rm p } = 2 T_{\rm e}$.

At $k_\perp \rho_{\rm p} \simeq 1$, AW/KAW turbulence has a range of
$k_\parallel$ values. However, we approximate the linear proton and electron
damping rates by assigning a single effective $|k_\parallel|$ to the
spectrum at~$k_\perp \rho_{\rm p} = 1$ given by the critical-balance condition~\citep{higdon84,goldreich95,cho04b,boldyrev06}
\begin{equation}
|k_\parallel| v_{\rm A}  = t_{\rm c}^{-1},
\label{eq:critbal} 
\end{equation} 
where
\begin{equation}
t_{\rm c} = \frac{\rho \delta v_{\rm p}^2}{Q}
\label{eq:tc} 
\end{equation} 
is the energy cascade time at $k_{\perp} \rho_{\rm p} = 1$, and
$\delta v _{\rm p}$ is the rms amplitude of AW/KAW velocity
fluctuations at $k_\perp \rho_{\rm p} \sim 1$. In writing
Equation~(\ref{eq:tc}), we have taken the total fluctuation energy per
unit volume at~$k_\perp \rho_{\rm p} \sim 1$ to be twice the kinetic
energy density~$\rho \delta v_{\rm p}^2/2$, and we have assumed that
dissipation at $k_\perp < \rho_{\rm p}^{-1}$ does not reduce the
cascade power at $k_\perp \rho_{\rm p} \sim 1$ much below the level
that is present throughout the inertial range.  There is some evidence
that the magnetic fluctuations in the solar wind are consistent with
turbulence theories based on critical
balance~\citep{horbury08,podesta09c,forman11}. However, there are
conflicting claims in the literature over the validity of
Equation~(\ref{eq:critbal}) when $z^+_{\rm rms} \gg z^-_{\rm rms}$, a
point to which we return in Section~\ref{sec:division}.

We assume that for length scales~$\lambda$ between $\rho_{\rm p}$ and
the perpendicular AW correlation length (outer scale)~$L_{\perp}$, the
rms amplitude of the AW velocity fluctuations at perpendicular
scale~$\lambda$ is~$\propto \lambda^{1/4}$ as in observations at $r=1
$~AU~\citep{podesta07,chen11},  direct numerical simulations of AW
turbulence in the presence of a strong background magnetic
field~\citep{maron01,muller05,mason06,perez08a}, and recent theories of strong
AW turbulence~\citep{boldyrev06,perez09a}. We thus take
\begin{equation}
\delta v_{\rm p} = \frac{z^+_{\rm rms}}{2} \left(\frac{\rho_{\rm
      p}}{L_\perp}\right)^{1/4}.
\label{eq:dvp} 
\end{equation} 
We set
\begin{equation}
L_\perp = L_{\perp \sun} \sqrt{\frac{a}{a_{\sun}}},
\label{eq:Lperp} 
\end{equation} 
where $L_{\perp \sun}$ is the value of $L_\perp$ at the coronal base,
so that $L_\perp$ increases in proportion to the cross-sectional
radius of the flux tube in our model. 

In addition to linear damping, KAWs at $k_\perp \rho_{\rm p} \sim 1$
undergo nonlinear damping through the ``stochastic heating'' of
protons~\citep{mcchesney87,chen01,johnson01}.  In stochastic proton
heating, AW/KAW fluctuations at $k_\perp \rho_{\rm p} \sim 1$ cause
proton orbits in the plane perpendicular to~$\bm{B}_0$ to become
stochastic, and the protons are subsequently energized by the
time-varying electrostatic potential.  Using numerical simulations of
test particles interacting with a spectrum of randomly phased AWs and
KAWs, \cite{chandran10a} found that
the effective damping rate of KAWs at $k_\perp \rho_{\rm p} \sim 1$
and $\beta_{\parallel \rm p} \lesssim 1$ 
from stochastic proton heating is
\begin{equation}
\gamma_{\rm s} = 0.18 \epsilon_{\rm p} \Omega_{\rm p} \,\exp\left(-\,\frac{c_2}{\epsilon_{\rm p}}\right),
\label{eq:gammas} 
\end{equation} 
where $c_2$ is a dimensionless constant, 
\begin{equation}
\epsilon_{\rm p} = \frac{\delta v_{\rm p}}{v_{\perp \rm p}},
\label{eq:epsp} 
\end{equation} 
and $v_{\perp \rm p} = \sqrt{2 k_{\rm B} T_{\perp \rm p}/m_{\rm p}}$ is
the proton perpendicular thermal speed. \cite{chandran10a} found that~$c_2
= 0.34$ for randomly phased AWs and KAWs, but conjectured that $c_2$
is smaller (and hence stochastic heating is more effective) in strong
AW/KAW turbulence, because much of the dissipation in strong AW/KAW
turbulence occurs within coherent structures in which the fluctuation
amplitudes are larger than their rms values \citep{dmitruk04}. In the
vicinity of such structures, proton orbits are more stochastic than on
average, allowing for more efficient stochastic
heating. \cite{chandran10b} developed a model of ion temperatures in
coronal holes based on stochastic heating, and this model matched the
observed ${\rm O}^{+5}$ temperature profile when $c_2$ was set equal
to~0.15.  In our model, we leave~$c_2$ as a free parameter.

The total effective damping rate of KAWs at $k_\perp \rho_{\rm p} =1 $~is 
\begin{equation}
\gamma_{\rm tot} = \gamma_{\rm e} + \gamma_{\rm p} + \gamma_{\rm s}.
\label{eq:gammatot} 
\end{equation} 
We define $\Gamma$ to be the fraction of the cascade power that is
dissipated at $k_\perp \rho_{\rm p} \sim 1$. This fraction is roughly
$\gamma_{\rm tot} t_{\rm c}$ when $\gamma_{\rm tot} t_{\rm c} \ll 1$
and roughly~1 when $\gamma_{\rm tot} t_{\rm c} \gg 1$. To interpolate
smoothly between these limits, we set
\begin{equation}
\Gamma = \frac{\gamma_{\rm tot} t_{\rm c}}{1 +
\gamma_{\rm tot} t_{\rm c}}.
\label{eq:Gamma} 
\end{equation}
Landau damping and transit-time damping of KAWs on protons contribute
to~$Q_{\parallel \rm p}$ but not to~$Q_{\perp \rm
  p}$~\citep{stix92}. On the other hand, stochastic heating leads
primarily to perpendicular proton heating when $\beta_{\parallel \rm
  p} \ll 1$ \citep{chandran10a}. \cite{johnson01} have shown that
stochastic heating leads to significant perpendicular proton heating
even at~$\beta_{\parallel \rm p} \sim 1$, and for simplicity we take
stochastic heating to contribute only to~$Q_{\perp \rm p}$, regardless
of the value of~$\beta_{\parallel \rm p}$. We divide the cascade power
that is dissipated at $k_\perp \rho_{\rm p} \sim 1$ into three parts
--- electron heating, perpendicular proton heating, and parallel
proton heating --- in proportion to the corresponding damping rates,
$\gamma_{\rm e}$, $\gamma_{\rm s}$, and $\gamma_{\rm p}$.  As
mentioned previously, we assume that the fraction of the cascade power
that is not dissipated at $k_\perp \rho_{\rm p} \sim 1$ cascades to
scales $\ll \rho_{\rm p}$, dissipates via interactions with electrons,
and contributes to~$Q_{\rm e}$. This procedure leads to
the relations 
\begin{equation}
Q_{\rm e} = \frac{(1 + \gamma_{\rm e} t_{\rm c}) Q}{1 + \gamma_{\rm tot} t_{\rm c}},
\label{eq:Qe} 
\end{equation} 
\begin{equation}
Q_{\perp \rm p} = \frac{\gamma_{\rm s} t_{\rm c}Q }{1 + \gamma_{\rm tot} t_{\rm c}},
\label{eq:Qperpp} 
\end{equation} 
and
\begin{equation}
Q_{\parallel \rm p} = \frac{\gamma_{\rm p} t_{\rm c}Q }{1 + \gamma_{\rm tot} t_{\rm c}}.
\label{eq:Qparp} 
\end{equation}

\section{Numerical Method}
\label{sec:method} 

\begin{figure*}[t]
\centerline{
\includegraphics[width=5.9cm]{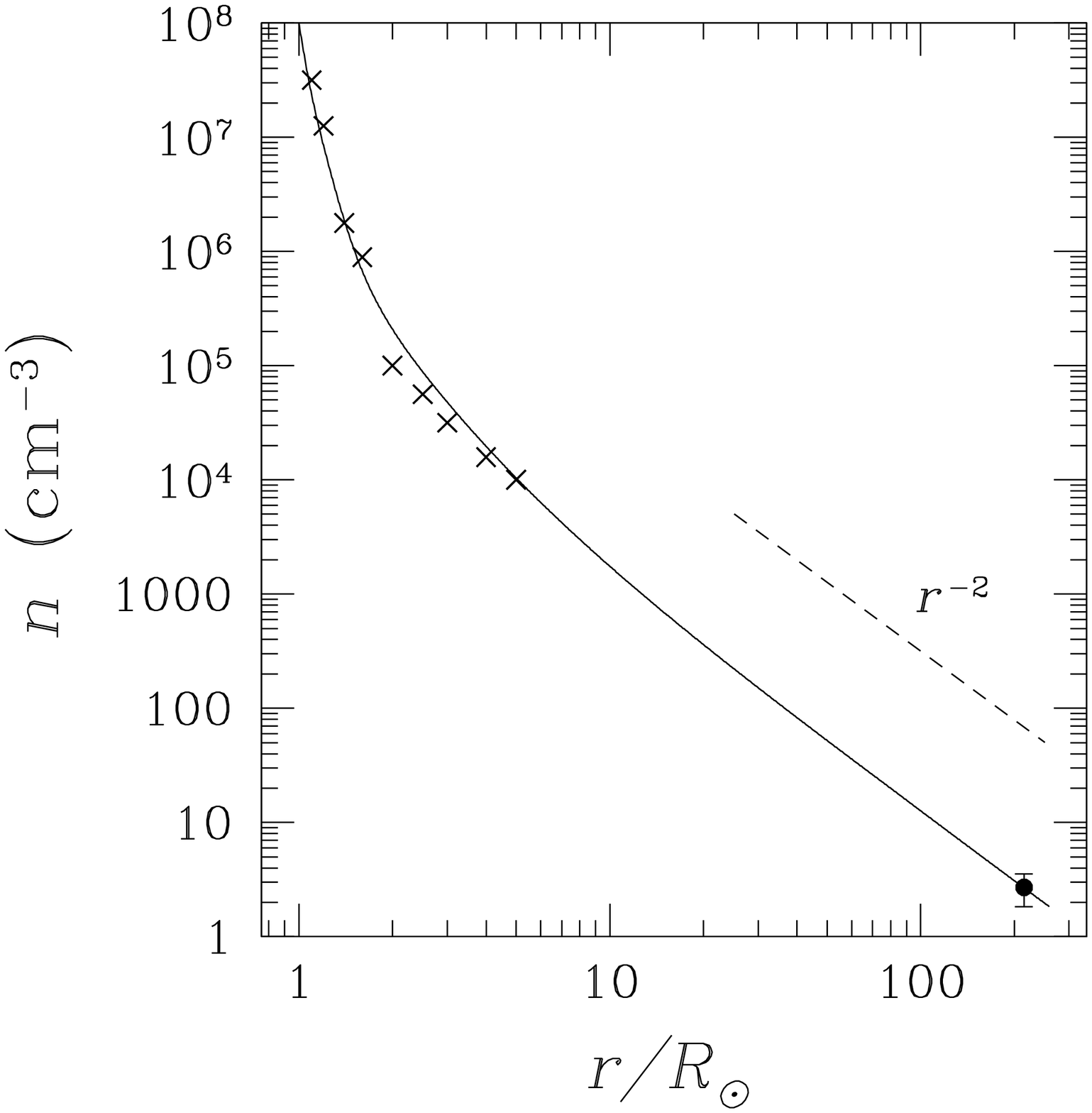}
\hspace{0.2cm} 
\includegraphics[width=5.9cm]{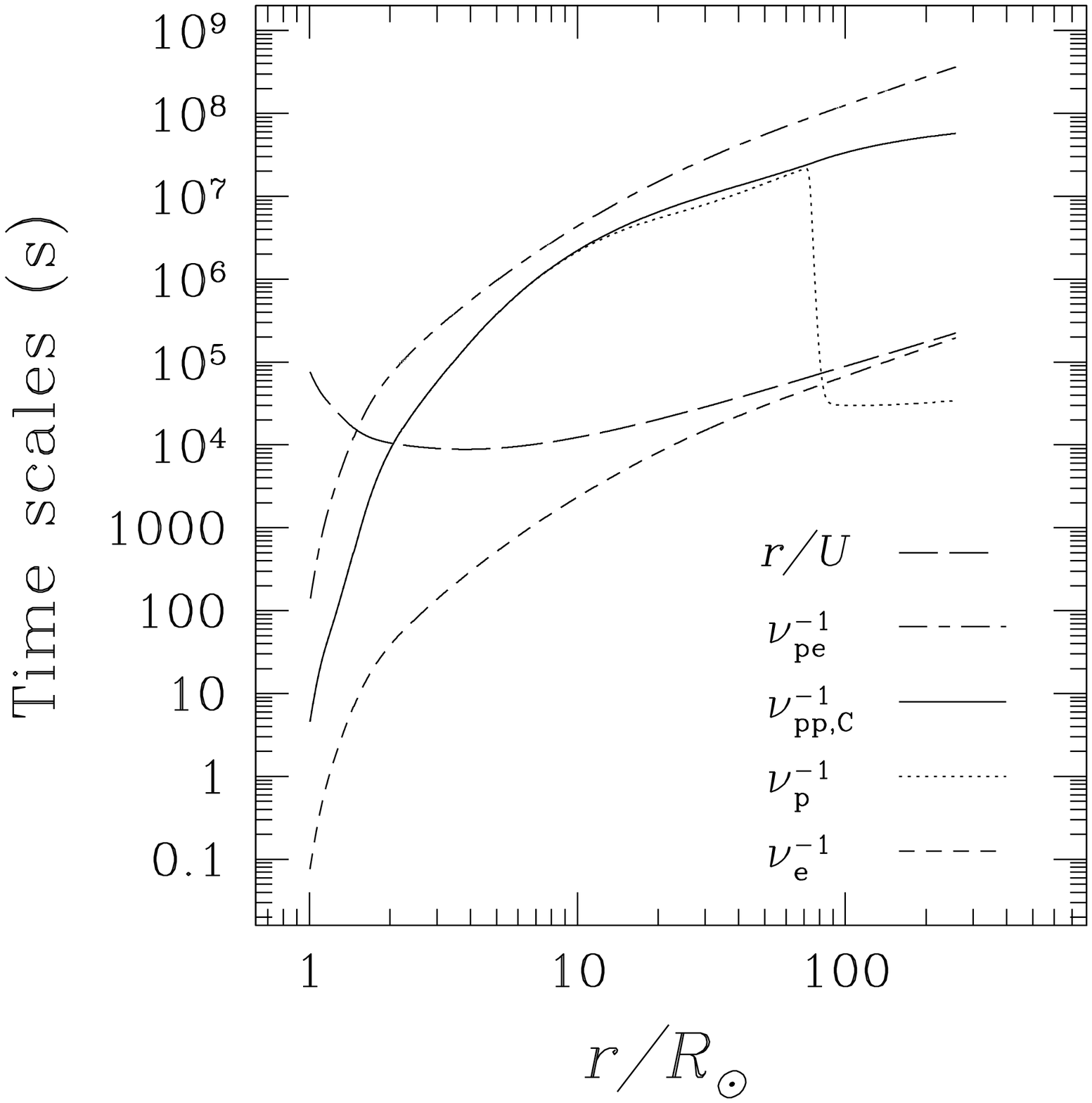}
\hspace{0.2cm} 
\includegraphics[width=5.9cm]{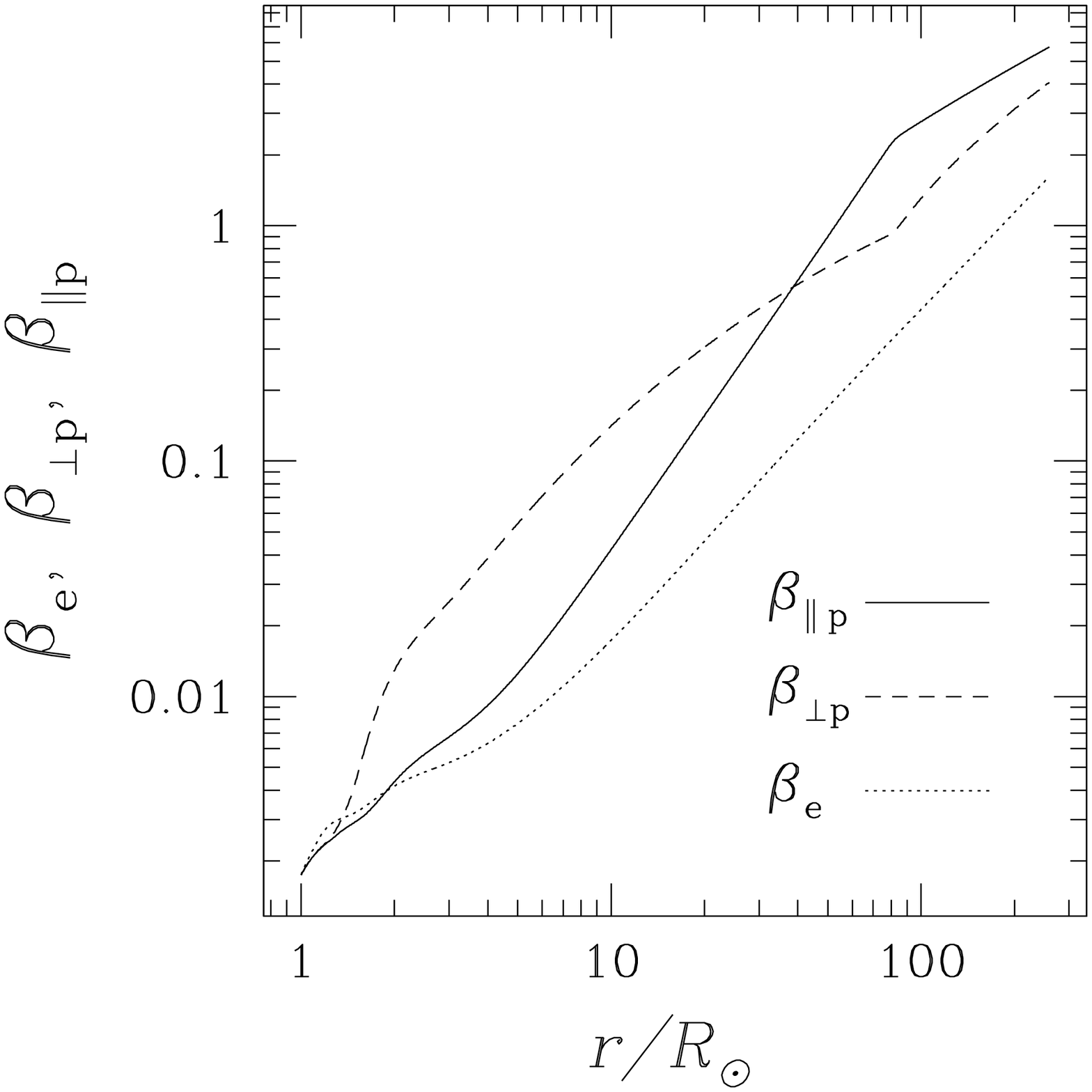}
}
\caption{{\em Left panel:} Proton (or electron) number density in
  model (solid line). The $\times$s show observed values in a polar
  coronal hole near solar minimum from Table 14.19 of
  \cite{allen73}. The filled circle is the mean proton density 
$n=2.7\pm 0.86\mbox{ cm}^{-3}$  at
  heliographic latitudes $> 36^\circ$ measured during Ulysses' first
  polar orbit, scaled to $r=1$~AU~\citep{mccomas00}. {\em
    Middle panel:} Important time scales in our steady-state numerical
  solution.  {\em Right panel:} The ratios of thermal to magnetic
  pressure: $\beta_{\parallel \rm p} = 8\pi n k_{\rm B} T_{\parallel
    \rm p}/B^2$, $\beta_{\perp \rm p} = 8 \pi n k_{\rm B} T_{\perp \rm
    p}/B^2$, and $\beta_{\rm e}= 8 \pi n k_{\rm B} T_{\rm e}/B^2$.
  \label{fig:swn}}
\end{figure*}

We integrate Equations~(\ref{eq:cont1}) through (\ref{eq:dEwdt})
forward in time until a steady state is reached using the implicit
numerical method described by \cite{hu97}. We use a logarithmic grid
in $r$ with grid points $r_i$ extending from $1 \;R_{\sun}$ to
$1.2$~AU, where $i= 0, 1, 2, \dots, N+1$ and~$N=1320$.  At each time
step, we update the variables by integrating the equations only at
$r_1, r_2, \dots, r_N$. We then update the variables at $r_0$ and
$r_{N+1}$ using the following boundary conditions. At $r=r_0$ we set
$n= n_{\sun}$, $T_e = T_{\perp \rm p} = T_{\parallel \rm p} =
T_{\sun}$, and ${\cal E}_{\rm w} = n_{\sun} m_{\rm p} (\delta
v_{\sun})^2$, where $n_{\sun}$, $T_{\sun}$, and $\delta v_{\sun}$ are
constants (see Table~\ref{tab:parameters}). We determine $U$,
$q_{\perp \rm p}$, and $q_{\parallel \rm p}$ at $r=r_0$ by linearly
extrapolating from the values at $r=r_1$ and $r=r_2$. We determine
$U(r_0)$ in this way rather than by fixing the value of~$U(r_0)$ so
that the variables can evolve towards a steady-state transonic
solution that passes smoothly through the sonic point. We determine
$q_{\perp \rm p}$ and $q_{\parallel \rm p}$ at $r=r_0$ by linear
extrapolation for the following reason. At radii slightly greater
than~$r_0$, the values of $q_{\perp \rm p}$ and $q_{\parallel \rm p}$
are determined to a good approximation by neglecting the terms on the
left-hand sides of Equations~(\ref{eq:dqperpdt}) and
(\ref{eq:dqpardt}), because the collision time~$\nu_{\rm p}^{-1}$ is
much shorter than the expansion time~$r/U$.  We call the values
of~$q_{\perp \rm p}$ and $q_{\parallel \rm p}$ determined in this way
the ``collisional values.'' If boundary conditions were imposed on
$q_{\perp \rm p}$ and $q_{\parallel \rm p}$ at $r=r_0$ that were
different from the collisional values, then in steady state a boundary
layer would develop at $r=r_0$ of thickness~$\sim U_0/\nu_{\rm p0}$,
where $U_0 = U(r_0)$ and $\nu_{\rm p0} = \nu_{\rm p}(r_0)$, and
$q_{\perp \rm p}$ and $q_{\parallel \rm p}$ would approach their
collisional values at $r\sim r_0 + U_0/\nu_{\rm p0}$. However, by
linearly extrapolating the values of $q_{\perp \rm p}$ and
$q_{\parallel \rm p}$ to $r=r_0$, we prevent such unphysical boundary
layers from appearing.  At $r=r_{N+1}$ we evaluate all variables by
linearly extrapolating from their values at $r=r_{N-1}$ and $r=r_{N}$.
For the solutions presented in Sections~\ref{sec:solution}
and~\ref{sec:swani_cyclotron} we used the following initial
conditions: $T_{\rm e} = T_{\perp \rm p} = T_{\parallel \rm p} =
T_{\sun} (3-2R_{\sun}/r) (r/R_{\sun})^{-2/7} $, $U = (655 \mbox{
  km/s})[1 + 20 (R_{\sun}/r)^{3}]^{-1}$, $n = n_{\sun} U_0 a_{\sun}/(U
a)$, ${\cal E}_{\rm w} = n m_{\rm p} (\delta v_{\sun})^2 $, and
$q_{\perp \rm p} = q_{\parallel \rm p} = 0$.

\section{A Numerical Example Resembling the Fast Solar Wind}
\label{sec:solution} 

In this section, we focus on a steady-state solution to
Equations~(\ref{eq:cont1}) through (\ref{eq:dEwdt}) in which the model
parameters are set equal to the values listed in Table~\ref{tab:parameters}.
Our choice for $L_{\perp \sun}$ is motivated by Faraday-rotation measurements along lines of sight
passing through the corona and near-Sun solar wind~\footnote{\cite{hollweg10}
  found that the magnetic fluctuations~$\delta B$ in the solar-wind
  model of~\cite{cranmer07} led to close agreement with the
  fluctuations in the Faraday rotation of radio transmissions from
  Helios near superior conjunction. A key parameter in the model of
  \cite{cranmer07} was~$L_\perp$, which was set equal to $75 [(1470
  \mbox{ Gauss})/B]^{1/2}$. For $B=11.8$~Gauss (the value of $B_{\sun}$
  in our numerical solutions), this leads to~$L_{\perp} =
  837$~km, which we have rounded to~$10^3$~km in choosing the value
  of~$L_{\perp \sun}$. It is possible that $L_{\perp \sun}$ is
  significantly larger than this value, but in this case $\delta B$
  would have to be significantly smaller than in the model of
  \cite{cranmer07} in order to be consistent with the Faraday rotation
  measurements, preventing AW turbulence from providing the heating
  needed to power the solar wind. We exclude the possibility that
  $L_{\perp \sun} \ll 10^3$~km, because then $\delta B$ would have to
  be much larger than in work of \cite{cranmer07} in order to be
  consistent with the Faraday rotation fluctuations, causing~$F_{\rm w}$ to
  be much greater than the total energy flux of the solar
  wind. \label{fn:faraday}}~\citep{hollweg10}.
After choosing the super-radial expansion factors $f_{\rm max}$
and~$R_1$, we determine $B_{\sun}$ from Equation~(\ref{eq:B})  and the
condition
\begin{equation}
B(\mbox{1 AU}) = 2.83 \mbox{nT},
\label{eq:B2} 
\end{equation} 
which is the mean radial magnetic field strength measured during
Ulysses' first polar orbit, scaled to $r=1 \mbox{
  AU}$~\citep{mccomas00}.  This yields $B_{\sun} = 11.8$~Gauss.  

\begin{table}[h]
\centering{
\caption{Parameters in Numerical Example
\label{tab:parameters} }
{\footnotesize 
\begin{tabular}{lcl}
\vspace{-0.2cm} 
&&  \\ 
\hline \hline 
\vspace{-0.25cm} 
& & \\
Quantity & & Value \\
\vspace{-0.25cm} 
&& \\
\hline
\vspace{-0.25cm} && \\
$n_{\sun}$ & $\dots\dots\dots\dots$ & $10^8 \mbox{ cm}^{-3}$ \\
$T_{\sun}$ & $\dots\dots\dots\dots$ & $7 \times 10^5\mbox{ K}$ \\
$\delta v_{\sun}$ & $\dots\dots\dots\dots$ & 41.4 km/s \\
$f_{\rm max}$& $\dots\dots\dots\dots$ & 9 \\
$R_1$ & $\dots\dots\dots\dots$ & $1.29 R_{\sun}$ \\
$L_{\perp \sun}$ & $\dots\dots\dots\dots$ & $10^3$ km \\
$c_{\rm d}$ & $\dots\dots\dots\dots$ & 0.75 \\
$c_{2}$ & $\dots\dots\dots\dots$ & 0.17 \\
$\alpha_{\rm H}$ & $\dots\dots\dots\dots$ & 0.75 \\
\vspace{-0.2cm} 
\\
\hline
\end{tabular}
\vspace{0.5cm} 
}
}
\end{table}

The resulting numerical solution resembles the fast solar wind in
several respects and illustrates the physical processes operating in
our model.  The density profile is shown in the left panel of
Figure~\ref{fig:swn}. Near the Sun, $n$~decreases rapidly with
increasing~$r$. At larger~$r$, as the wind approaches its asymptotic
speed, $n$~becomes roughly proportional to~$r^{-2}$. The Coulomb
collision time scales $\nu_{\rm e}^{-1}$, $\nu_{\rm pp,C}^{-1}$, and
$\nu_{\rm pe}^{-1}$ are shown in the middle panel of
Figure~\ref{fig:swn}. Also plotted are the advection time
scale~$t_{\rm adv} = r/U$ and the total proton scattering time scale
$\nu_{\rm p}^{-1}$, which includes the effects of
temperature-anisotropy instabilities.  Close to the coronal base, the
density is sufficiently large that Coulomb collisions play an
important role, acting to maintain $T_{\rm e} \simeq T_{\perp \rm p}
\simeq T_{\parallel \rm p}$. Farther from the Sun, however, Coulomb
collisions cause only a negligible amount of proton-electron energy
exchange and proton temperature isotropization. In the right panel of
Figure~\ref{fig:swn}, we plot the ratios of thermal to magnetic
pressure, $\beta_{\rm e}$, $\beta_{\perp \rm p}$, and
$\beta_{\parallel \rm p}$ (defined in the figure caption), which vary
from values $\ll 1$ near the Sun to values larger than~1 at 1~AU.

\begin{figure*}[t]
\centerline{
\includegraphics[width=5.9cm]{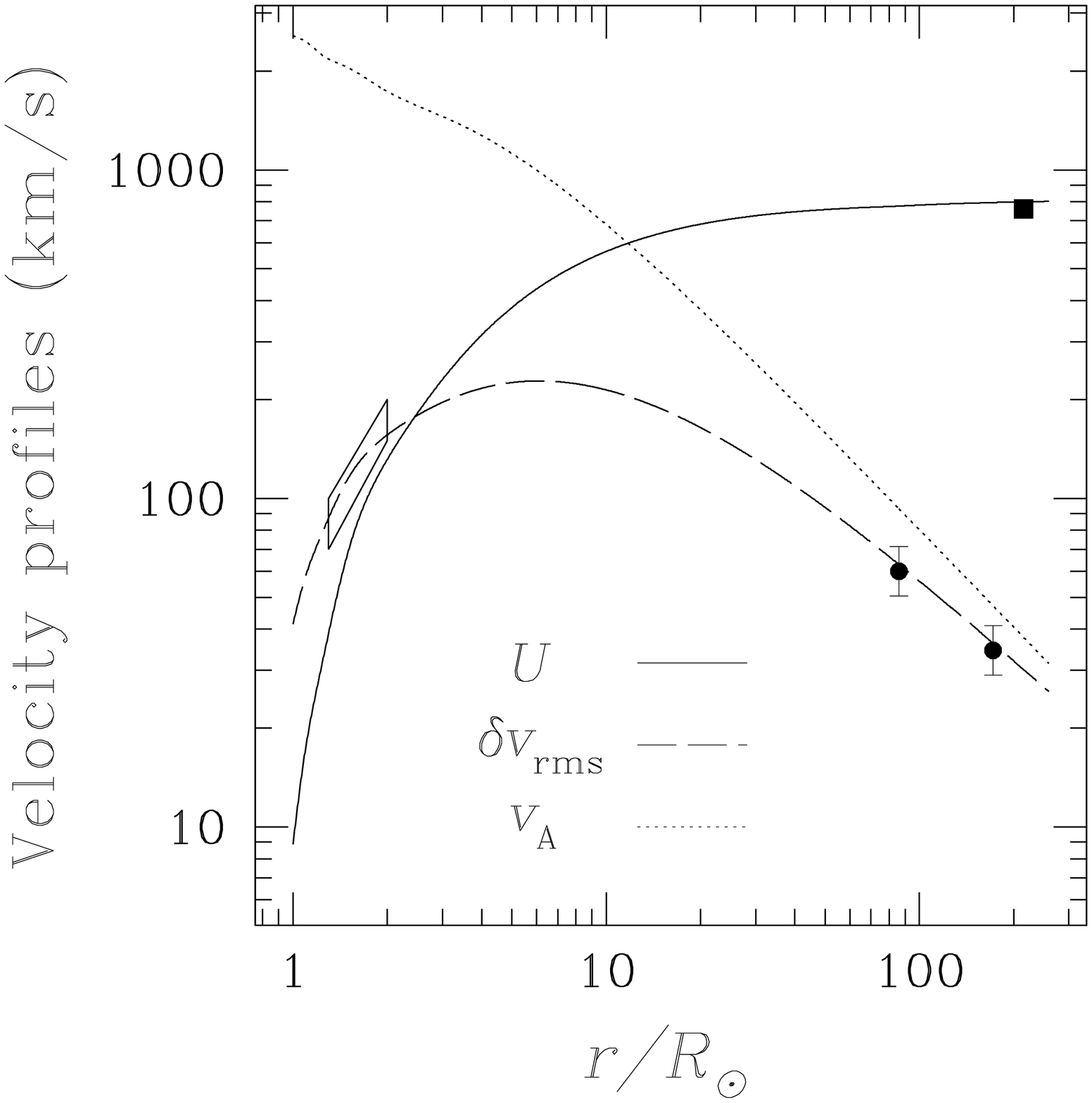}
\hspace{0.2cm} 
\includegraphics[width=5.9cm]{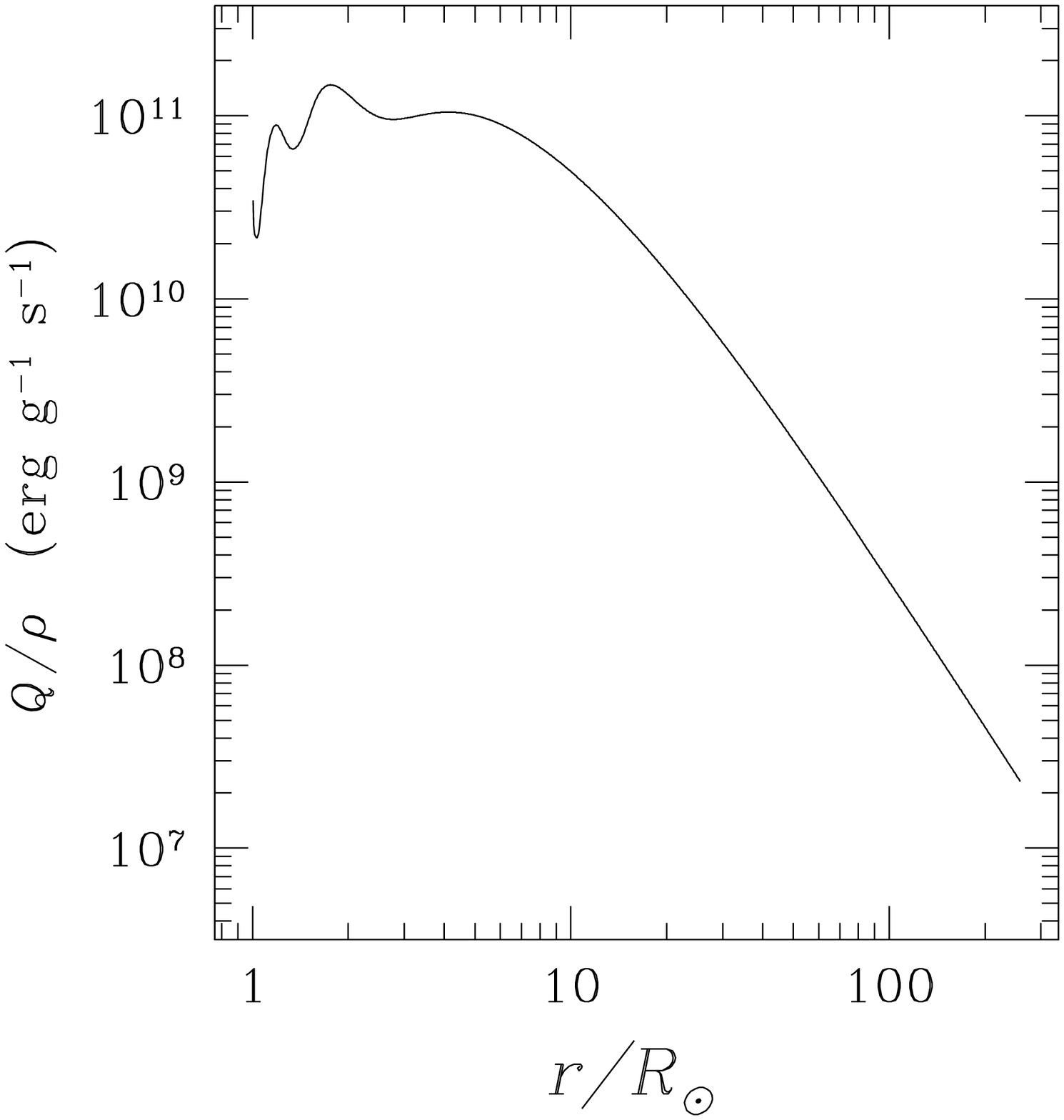}
\hspace{0.2cm} 
\includegraphics[width=5.9cm]{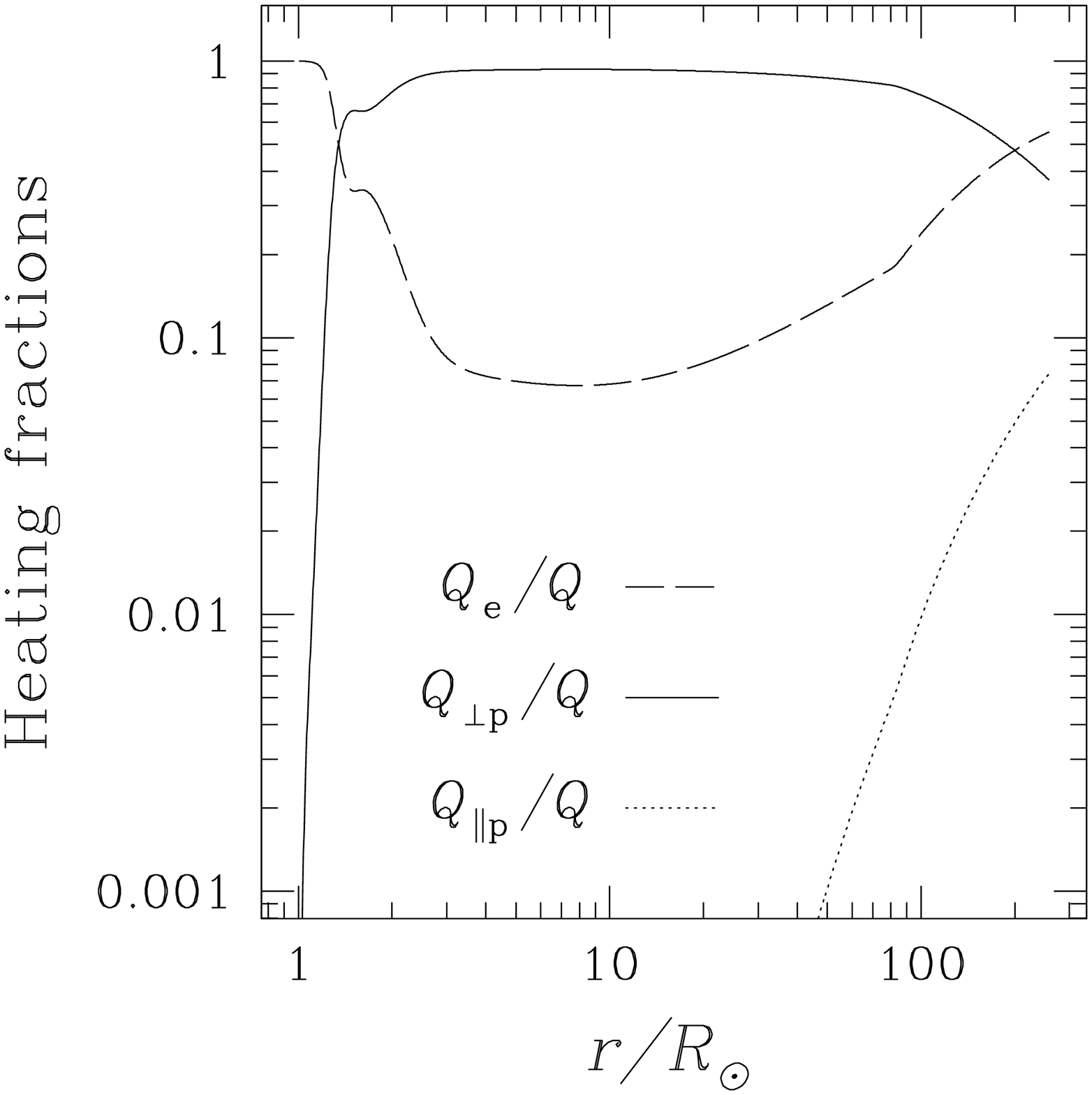}
}
\caption{{\em Left panel:} The open box
  represents the range of upper limits on $\delta v_{\rm rms}$
  obtained by \cite{esser99}. The filled circles are Helios measurements of $\delta
  v_{\rm rms}$ \citep{bavassano00}.
The filled square is the median proton flow speed $U = 761$~km/s at 
  heliographic latitudes $>36^\circ$ during Ulysses' first polar
  orbit, scaled to $r = \mbox{1 AU}$ 
  \citep{mccomas00}. The size of this square represents the range
 $702 \mbox{ km/s} < U < 803 \mbox{ km/s}$ corresponding to 
  the $5^{\rm th}$ through $95^{\rm th}$ percentiles of the measured
  distribution.
{\em Middle panel:} Total turbulent heating rate per unit mass.
{\em Right panel:} The fractions of the turbulent heating power that go
to electron heating, perpendicular proton heating, and parallel proton heating.
\label{fig:swdv} }
\end{figure*}

The profiles of the wind speed~$U$ and Alfv\'en speed~$v_{\rm A}$ 
are shown in the left panel of
Figure~\ref{fig:swdv}, along with the rms amplitude of the fluctuating
AW velocity
\begin{equation}
\delta v_{\rm rms} = \sqrt{\frac{{\cal E}_{\rm w}}{\rho}}.
\label{eq:dvrms} 
\end{equation} 
In writing Equation~(\ref{eq:dvrms}), we have ignored the energy
associated with AWs propagating towards the Sun in the solar-wind
frame, so that $\delta v_{\rm rms} = z^+_{\rm rms}/2$. In our
numerical solution, the Alfv\'en critical point occurs at 
\begin{equation}
r_{\rm A} = 11.7 R_{\sun},
\label{eq:rA} 
\end{equation} 
$U(r_{\rm A}) = 598$~km/s, and $U(1 \mbox{ AU}) = 800$~km/s. The AW
fluctuations lead to a turbulent heating rate per unit mass~$Q/\rho$
that is shown in the middle panel of Figure~\ref{fig:swdv}.

\begin{figure*}[t]
\centerline{
\includegraphics[width=5.9cm]{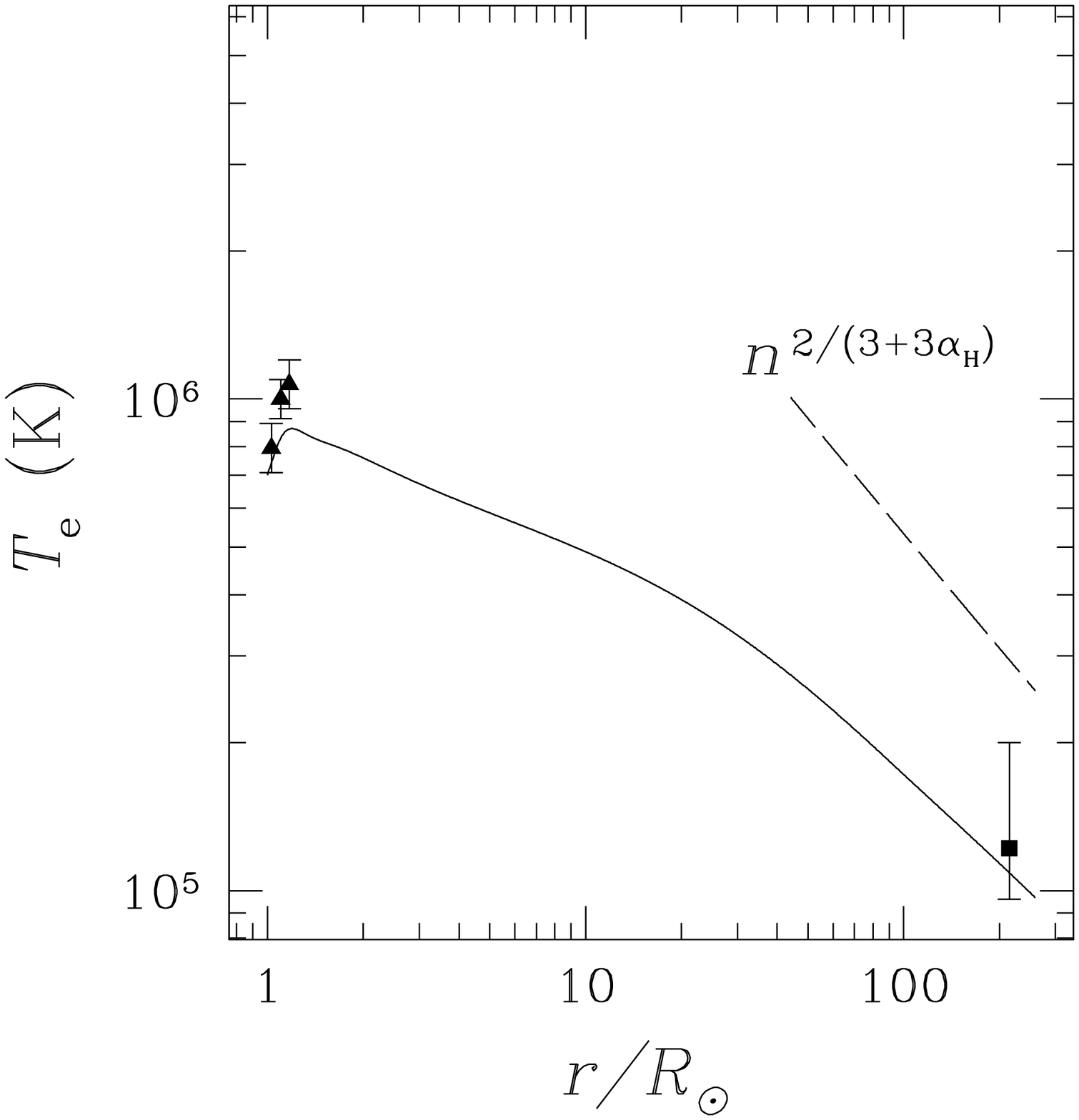}
\hspace{0.2cm} 
\includegraphics[width=5.9cm]{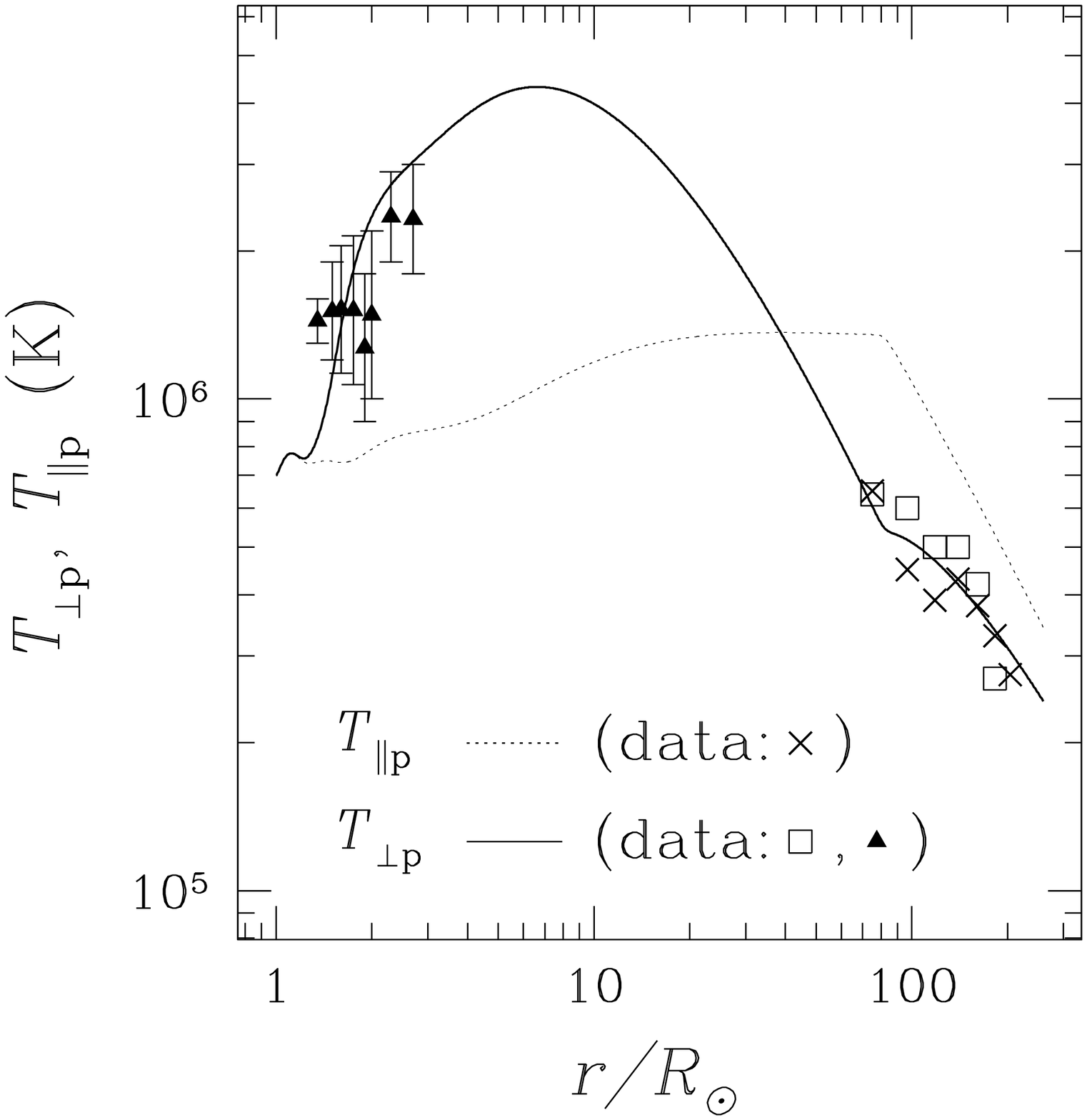}
\hspace{0.2cm} 
\includegraphics[width=5.9cm]{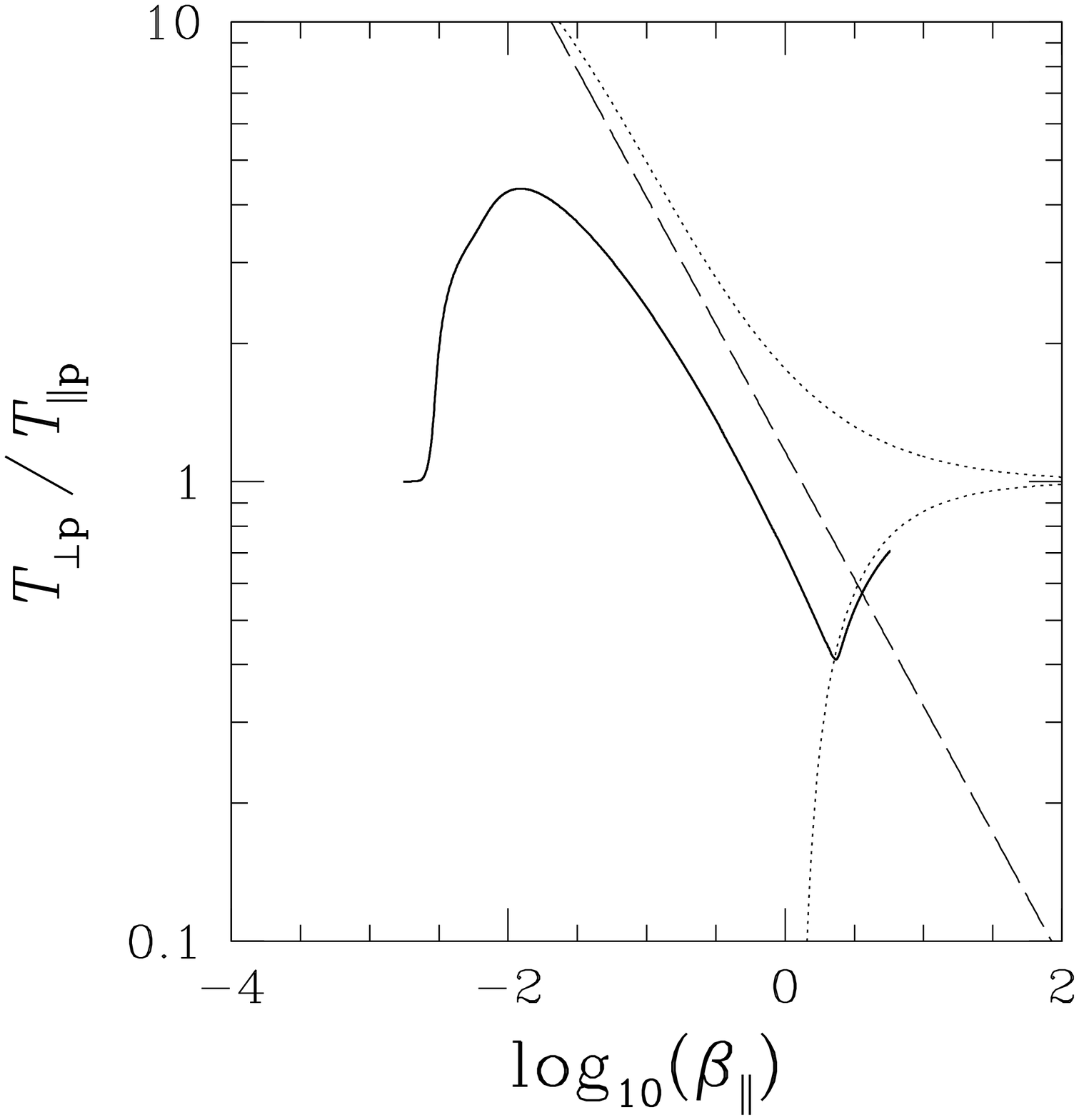}
}
\caption{{\em Left panel:} Triangles are electron temperatures
  inferred from spectroscopic
  observations of a polar coronal hole~\citep{landi08}. The square is
  the mean electron temperature in ISEE 3 and Ulysses measurements of
  fast-wind streams with $600 \mbox{ km/s} < U < 700 \mbox{ km/s}$ \citep{newbury98}.
{\em Middle panel:} Triangles are proton kinetic temperatures inferred
from  UVCS
measurements of a polar coronal hole \citep{esser99}. Open squares 
($\times$s) are perpendicular (parallel) proton temperatures measured by Helios in fast solar-wind streams with $700 \mbox{ km/s} < U < 800 \mbox{ km/s}$~\citep{marsch82b}.
{\em Right panel:} 
 The value of $T_{\perp \rm p}/T_{\parallel
    \rm p}$ in our numerical solution progresses from left to right
  along the solid-line curve as $r$ increases from $1 R_{\sun}$ to
  $258 R_{\sun}$. The dotted lines give the instability thresholds for
  the mirror instability (upper curve) and oblique
  firehose instability (lower curve) from \cite{hellinger06}, at which
  the maximum instability growth rates equal $ 10^{-3} \Omega_{\rm
    p}$, where $\Omega_{\rm p}$ is the proton cyclotron frequency. The
  long-dashed line is the fit $T_{\perp \rm p}/T_{\parallel \rm p} =
  1.16 \beta_{\parallel \rm p}^{-0.553}$ obtained by \cite{marsch04}
  to Helios data of high-speed wind streams between~$r=0.29$~AU and $r=0.98$~AU.
  \label{fig:swT} }
\end{figure*}

The partitioning of the turbulent heating power between $Q_{\rm e}$,
$Q_{\perp \rm p}$, and~$Q_{\parallel \rm p}$ is shown in the right
panel of Figure~\ref{fig:swdv}. At $r< 1.35 R_{\sun}$, electrons
absorb most of the dissipated AW/KAW energy because $\delta v_{\rm
  rms}$ and hence~$\epsilon_{\rm p}$ are comparatively small (making
stochastic heating weak) and because $\beta_{\parallel \rm p} \ll 1$
(making proton Landau damping and proton transit-time damping weak).
Between $r=1.35 R_{\sun}$ and $r=185 R_{\sun}$, stochastic heating is
the most efficient dissipation mechanism in the model, and $Q_{\perp
  \rm p}> 0.5Q$. At $r > 185 R_{\sun}$, $Q_{\rm e}$ is the largest of
the three heating rates. The parallel proton heating rate remains $<
Q_{\rm e}$ at large~$r$ despite the fact that $\beta_{\parallel \rm p}
\gtrsim 1$ and $\gamma_{\rm p} > \gamma_{\rm e}$. The reason for this
is that at large~$r$ a significant fraction of the turbulent energy
cascades to perpendicular scales~$\ll \rho_{\rm p}$ at which the
fluctuations dissipate on the electrons.

\begin{figure*}[t]
\centerline{
\includegraphics[width=5.9cm]{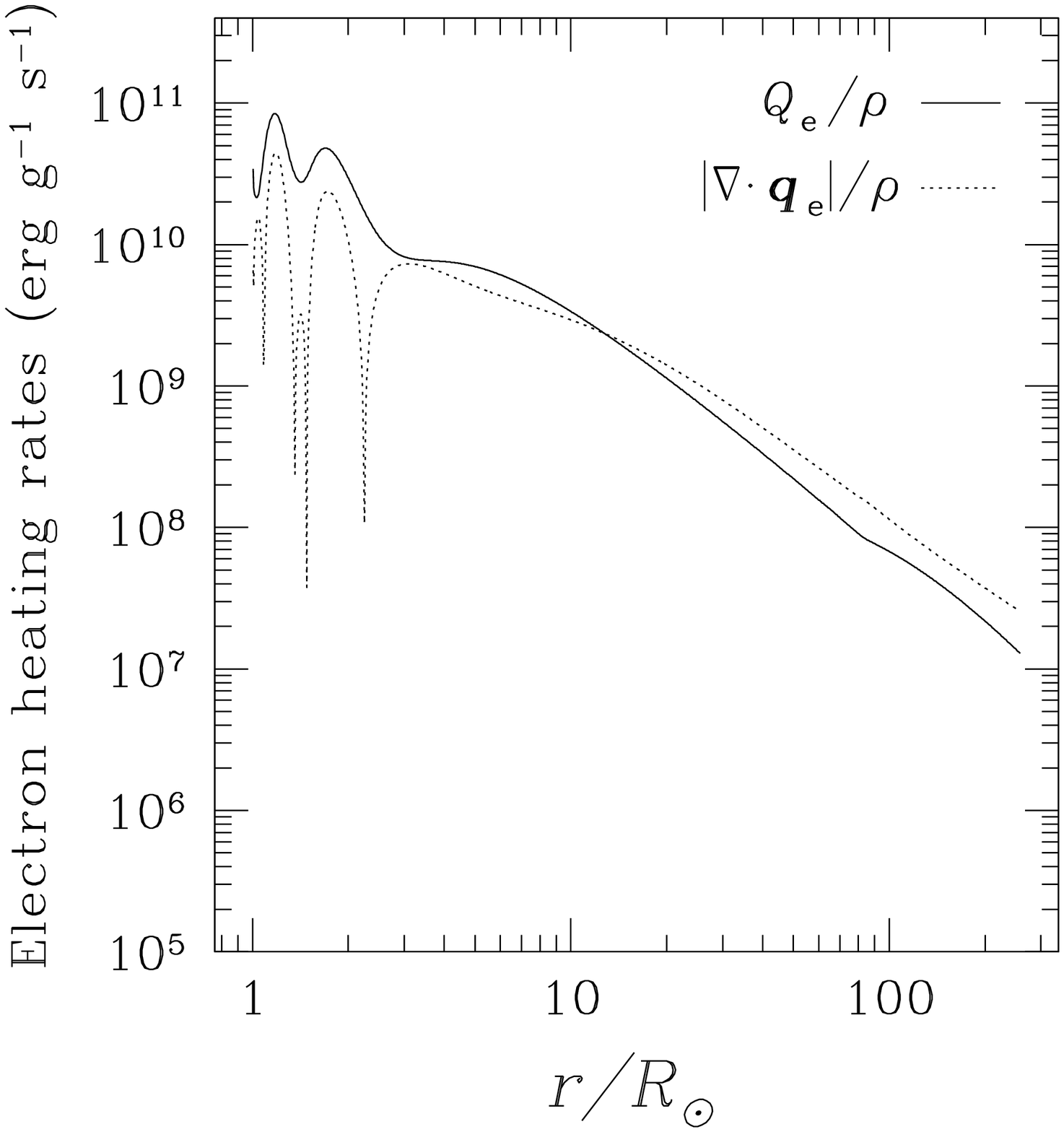}
\hspace{0.2cm} 
\includegraphics[width=5.9cm]{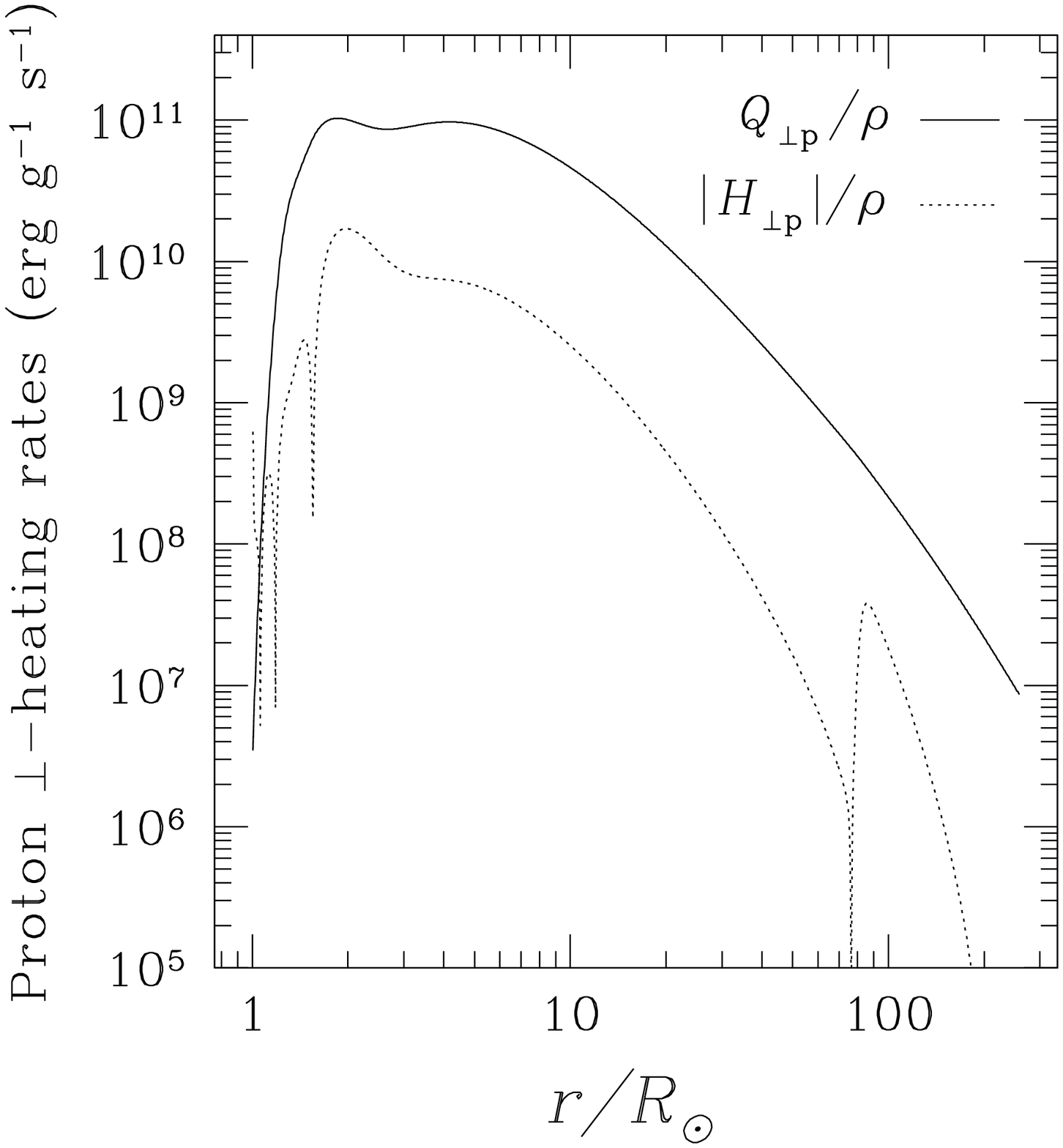}
\hspace{0.2cm} 
\includegraphics[width=5.9cm]{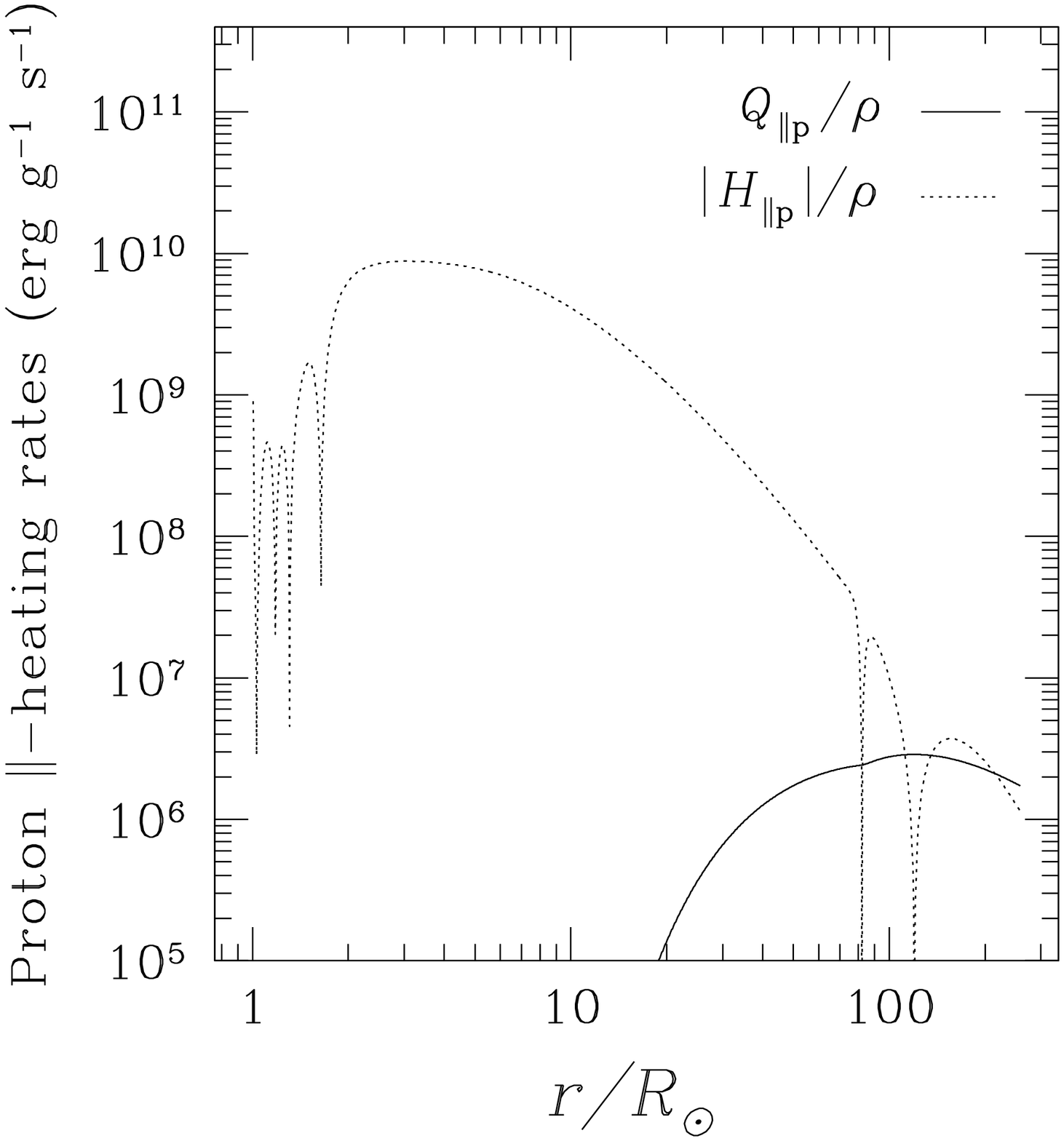}
}
\caption{Heating rates associated with the heat fluxes and the
  dissipation of AW/KAW turbulence.
\label{fig:swQe} }\vspace{0.5cm}
\end{figure*}

In Figure~\ref{fig:swT} we plot $T_{\rm e}$, $T_{\perp \rm p}$, and
$T_{\parallel \rm p}$, and in Figure~\ref{fig:swQe} we plot the
heating rates that help determine these temperature profiles.  Between
$R_{\sun}$ and $2 R_{\sun}$, there are ripples in the plots of $Q_{\rm
  e}$ and~$Q$, which result from local flattenings in the Alfv\'en
speed profile (see Equations~(\ref{eq:zminus}) and
(\ref{eq:Q})).\footnote{The oscillations in $Q$ and $Q_{\rm e}$
  between $r=R_{\sun}$ and $r\simeq 2 R_{\sun}$ result in part from
  our approximating $z^-_{\rm rms}$ and $Q$ in
  Equations~(\ref{eq:zminus}) and (\ref{eq:Q}) based on the local
  value of $|\partial v_{\rm A}/\partial r|$.  A more realistic
  treatment would account for the fact that $z^-$ AWs propagate some
  distance along the magnetic field before their energy cascades and
  dissipates, so that the local values of $z^-_{\rm rms}$ and~$Q$
  depend upon the value of $|\partial v_{\rm A}/\partial r|$
  throughout some range of radii.} The ripples in $Q_{\rm e}$ cause
$T_{\rm e}$ and $q_{\rm e}$ to vary in such a way that the electron
conductive heating rate partially offsets the variations in $Q_{\rm
  e}$. This can be seen in left panel of Figure~\ref{fig:swQe}, in
which the electron conductive heating rate $-\bm{\nabla} \cdot
\bm{q}_{\rm e}$ is positive at $r>2.25R_{\sun}$ but alternates sign at
each sharp dip in the plot of~$|\bm{\nabla} \cdot \bm{q}_{\rm e}|$. At
$R_{\sun} < r \lesssim 2.5 R_{\sun}$, AW turbulence is the dominant
heat source for electrons, as the electron heat flux acts primarily to
cool the electrons in this region.  The left panel of
Figure~\ref{fig:swQe} indicates that a significant fraction of the
turbulent heating power deposited into the electrons between
$R_{\sun}$ and $2.5 R_{\sun}$ is conducted away to either larger or
smaller radii. At $r\gtrsim 3 R_{\sun}$, $Q_{\rm e}$ and $|\bm{\nabla
} \cdot \bm{q}_{\rm e}|$ are of similar magnitude, although
$|\bm{\nabla } \cdot \bm{q}_{\rm e}|/Q_{\rm e}$ grows to a value~$\sim
2$ as $r$ increases to~$\mbox{1.2 AU}$. At $r \gtrsim 100 R_{\sun}$
the electrons approach the state described by \cite{hollweg76}, in
which collisionless heat flux is the only source of electron heating
and
\begin{equation}
T_{\rm e} \propto n^{2/[3(1+\alpha_{\rm H})]}.
\label{eq:TH76} 
\end{equation} 
However, the $T_{\rm e}$ profile remains slightly flatter than the
scaling in Equation~(\ref{eq:TH76}) because of turbulent heating.

At $2 R_{\sun} \lesssim r < 71 R_{\sun}$, $T_{\perp \rm p}$ is
determined by a balance between turbulent heating and solar-wind
expansion, with the proton heat flux and collisions playing only a
minor role, as shown in the middle panels of Figures~\ref{fig:swn}
and~\ref{fig:swQe}. The quantity $H_{\perp \rm p}$ plotted in
Figure~\ref{fig:swQe} is the perpendicular heating rate resulting from
the proton heat flux,
\begin{equation}
H_{\perp \rm p } = - \frac{1}{a^2}\frac{\partial }{\partial r}\left(a^2
  q_{\perp \rm p}\right).
\label{eq:Hperp} 
\end{equation} 
The $T_{\perp \rm p}$ profile at $2 R_{\sun} \lesssim r < 71 R_{\sun}$
is also affected by the self-limiting nature of stochastic heating. As
$T_{\perp \rm p}$ increases, the fluctuations in the electrostatic
potential energy at the proton-gyroradius scale become a smaller
fraction of the average perpendicular kinetic energy per
proton,~$k_{\rm B} T_{\perp \rm p}$. As a consequence, these
fluctuations have less effect on the proton gyro-motion, and the
proton orbits become less stochastic. This leads to a strong reduction
in the stochastic heating rate when $T_{\perp \rm p}$ exceeds a
certain threshold that depends upon $\delta v_{\rm p}$ and~$c_2$, as
described in more detail by \cite{chandran10b}.  At $r= 71 R_{\sun}$,
the plasma encounters the threshold of the oblique firehose
instability, causing $\nu_{\rm p}$ to increase abruptly and leading to
sharp cusps in the $T_{\perp \rm p}$ and $T_{\parallel \rm p}$
profiles plotted in Figure~\ref{fig:swT}. At $r> 71 R_{\sun}$ the
temperature anisotropy ratio~$T_{\perp \rm p}/T_{\parallel \rm p}$
evolves approximately along the oblique-firehose instability
threshold, as shown in the right panel of Figure~\ref{fig:swT}.

The parallel proton heating rate associated with the proton heat flux,
\begin{equation}
H_{\parallel \rm p} 
= - \frac{1}{a} \frac{\partial }{\partial r} ( a q_{\parallel \rm p})
+ \frac{q_{\perp \rm p}}{a}\,\frac{\partial a}{\partial r},
\label{eq:Hpar} 
\end{equation} 
is shown in the right panel of Figure~\ref{fig:swQe}, along
with~$Q_{\parallel \rm p}$.  Although $|H_{\parallel \rm p}| $ is
small compared to~$Q$, it is larger than $Q_{\parallel \rm p}$ at $
r \lesssim 70 R_{\sun}$. Despite the fact that $|H_{\parallel \rm p}| \ll Q$, the proton
heat flux causes $T_{\parallel \rm p}$ to increase with increasing~$r$
within this range of radii. This is possible because
solar-wind expansion has only a small effect on $T_{\parallel \rm p}$
at these radii. As can be seen from
Equation~(\ref{eq:dTpardt}), solar-wind expansion acts to make
$T_{\parallel \rm p} \propto n^2/B^2$ in the absence of competing
effects.  As the solar wind approaches its asymptotic speed, $n$
becomes approximately proportional to~$r^{-2}$. Likewise,
when $r$ exceeds a few~$R_{\sun}$, $B\propto r^{-2}$, at least in our
model in which solar rotation is neglected (see Section~\ref{sec:rotation}).
When both $n$ and $B$ are~$\propto r^{-2}$, 
(double) adiabatic expansion neither increases nor decreases~$T_{\parallel \rm p}$. 

\begin{figure*}[t]
\centerline{
\includegraphics[width=5.9cm]{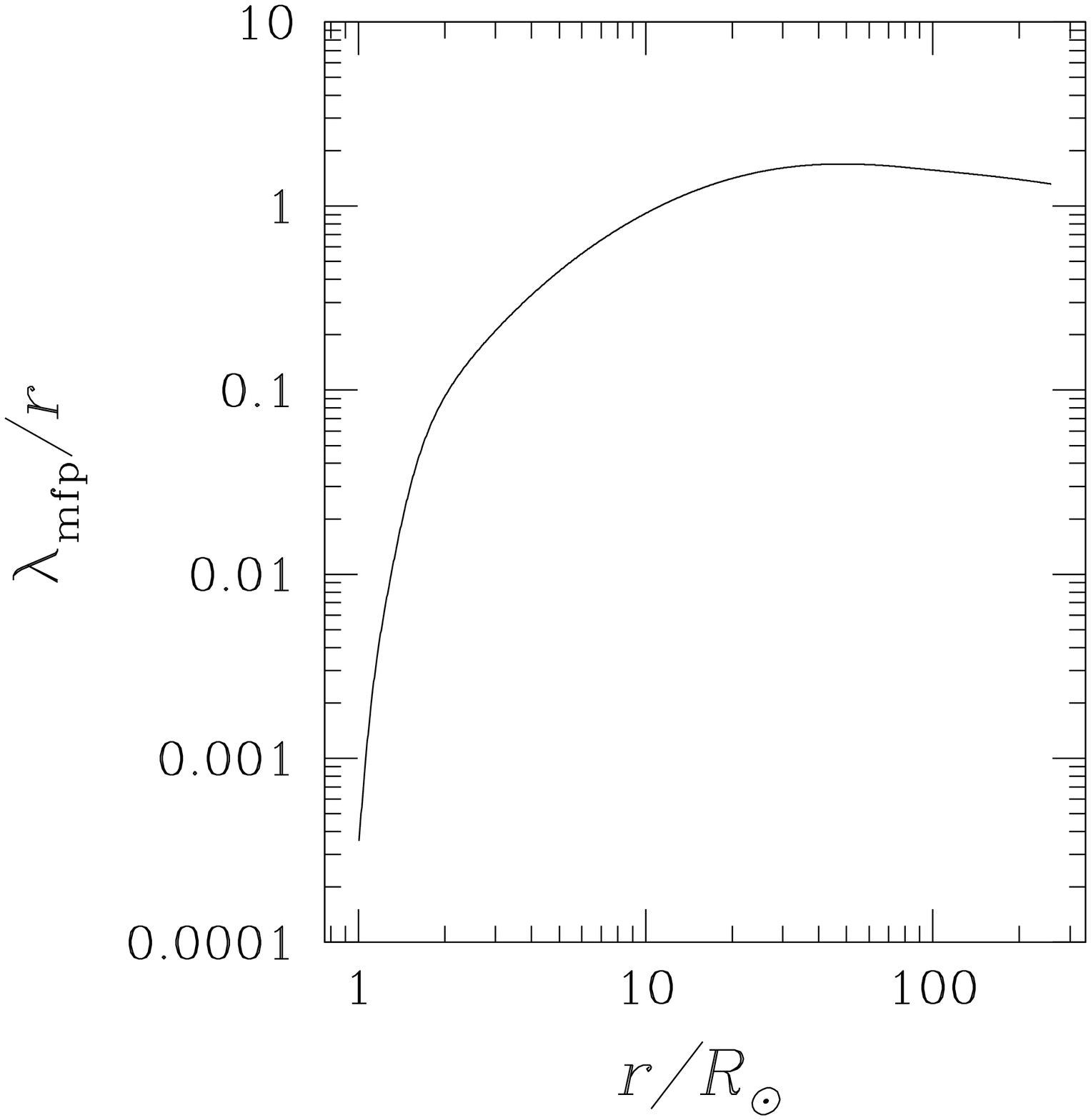}
\hspace{0.2cm} 
\includegraphics[width=5.9cm]{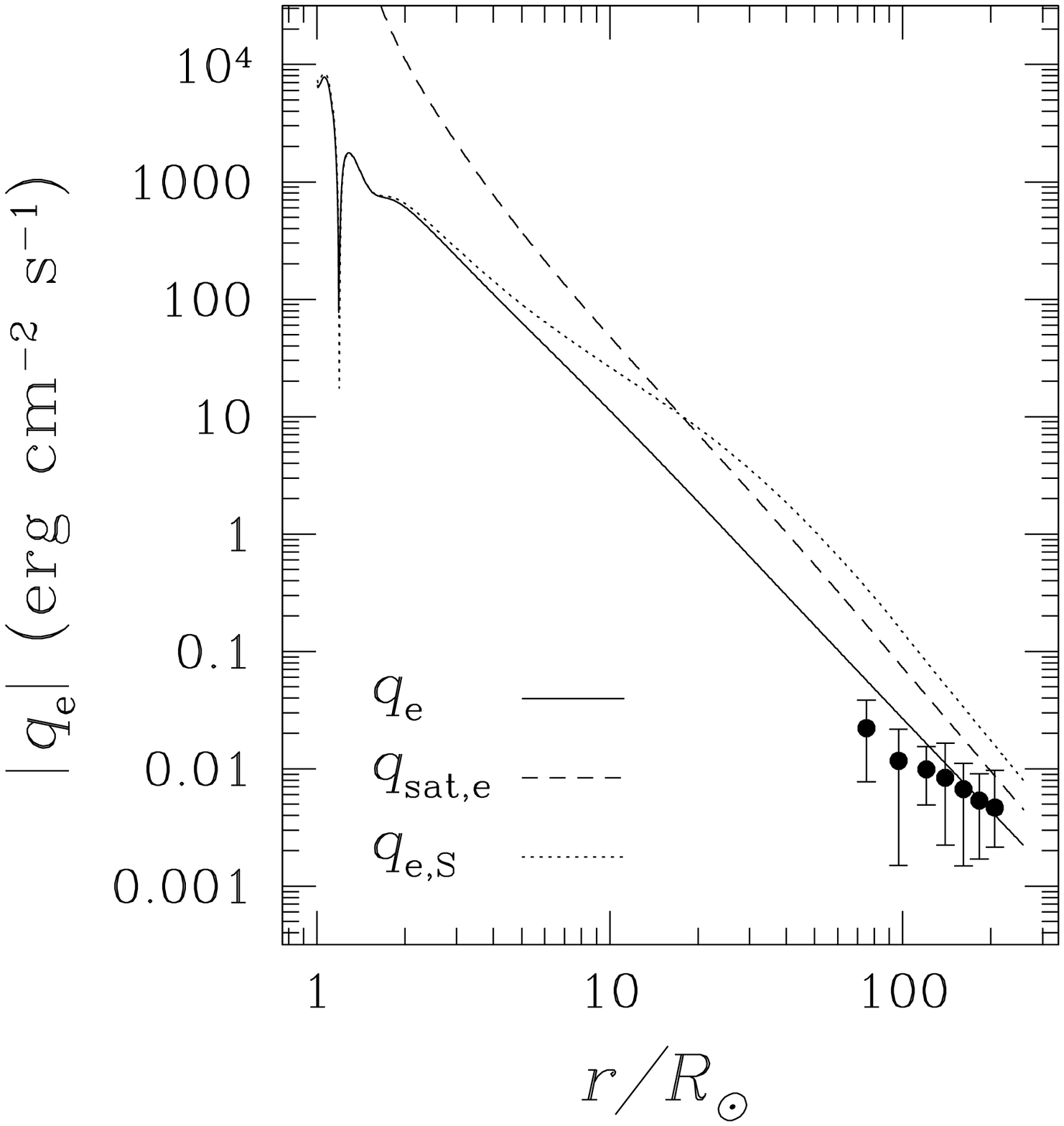}
\hspace{0.2cm} 
\includegraphics[width=5.9cm]{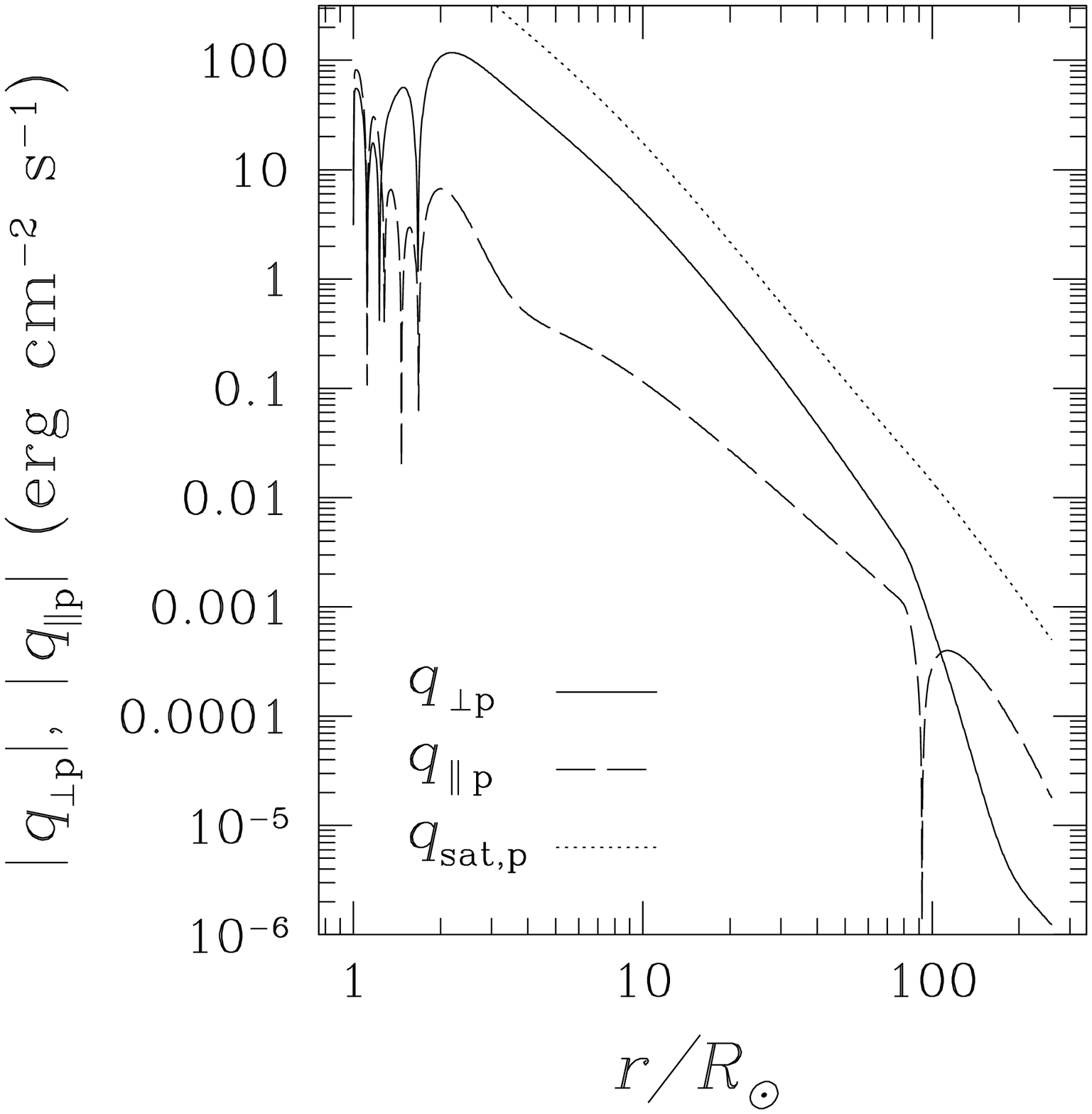}
}
\caption{{\em Left panel:} $\lambda_{\rm mfp}$ is the electron
Coulomb mean free path.
{\em Middle panel:} The electron heat flux~$q_{\rm e}$, free-streaming heat flux
  $q_{\rm sat,e}$, and Spitzer-H\"arm heat
  flux~$q_{\rm e, S}$. The circles
  are Helios measurements of the electron heat flux in high-speed wind
  with~$U> 600$~km/s \citep{marsch84}. 
{\em Right panel:} The proton heat fluxes~$ q_{\perp \rm p}$ and $q_{\parallel \rm p}$ and free-streaming heat flux 
  $q_{\rm sat,p}$.
\label{fig:swq} }\vspace{0.5cm}
\end{figure*}

The electron Coulomb mean free path~$\lambda_{\rm mfp}$ and electron
heat fluxes $q_{\rm e}$ are plotted in the left and middle panels of
Figure~\ref{fig:swq}. The electron heat flux transitions from the
collisional regime to the collisionless regime in our model at $r= r_{\rm H}
= 5 R_{\sun}$, approximately the point at which $\lambda_{\rm mfp} =
r/2$ as in the collisionless-heat-flux model of \cite{hollweg74a,hollweg76}.  In
the collisionless regime, $q_{\rm e}$ is smaller than
the Spitzer-H\"arm heat flux~$q_{\rm e, S}$ and somewhat smaller than the free-streaming
heat flux
\begin{equation}
q_{\rm sat,e} = 1.5 n k_{\rm B} T_{\rm e} v_{\rm
  te},
\label{eq:qsate} 
\end{equation} 
but comparable to Helios measurements of the electron heat flux in the
fast solar wind~\citep{marsch84}.

The proton heat fluxes are plotted in the right panel of
Figure~\ref{fig:swq}. As this figure shows, $q_{\perp \rm p}$ and
$q_{\parallel \rm p}$ are significantly smaller than the
free-streaming heat flux
\begin{equation}
q_{\rm sat,p} = 1.5 n k_{\rm B} T_{\rm p} v_{\rm tp},
\label{eq:qsatp} 
\end{equation} 
where $v_{\rm tp } = \sqrt{k_{\rm B} T_{\rm p}/m_{\rm p}}$. This can
be understood on a qualitative level from the following argument. If a
proton temperature gradient were set up in a collisionless plasma with
no background flow and no initial heat flux, then the proton heat flux
would grow in time, approaching a level comparable to the
free-streaming value after a time $\sim t_{\rm cross} = l_{\rm
  T}/v_{\rm tp}$, where~$l_{\rm T} = T_{\rm p}/|\bm{\nabla} T_{\rm
  p}|$. In our model, $l_{\rm T} \sim r$, and this ``crossing time
scale'' is a factor of $\sim M$ larger than the expansion time scale
of the solar wind,~$t_{\rm adv} = r/U$, where $M = U/v_{\rm tp}$ is
the Mach number. In our numerical solution, $M$ equals 3.56 at $r=10
R_{\sun}$ and grows monotonically with increasing~$r$ to a value
of~15.1 at $r= 215 R_{\sun}$. Thus, throughout most of our solution,
$t_{\rm adv} = r /U \ll t_{\rm cross}$. As a result, the protons in
our model do not have time to set up a heat flux comparable to the
free-streaming heat flux within the time it takes for the plasma to
double its distance from the Sun, which reduces $q_{\perp \rm p}$ and
$q_{\parallel \rm p}$ relative to their values in a stationary plasma
with comparable density and temperature profiles.

\begin{figure}[h]
\centerline{
\includegraphics[width=7cm]{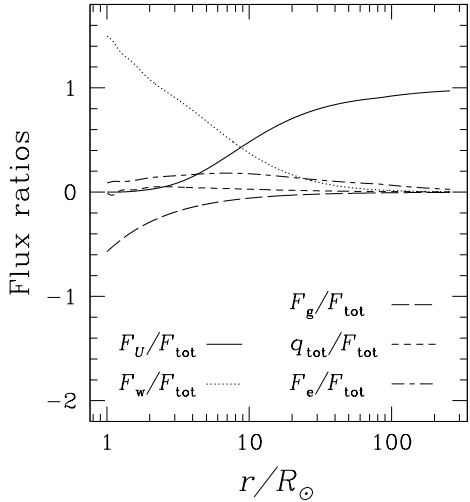}
}
\caption{The fractions of the total energy flux~$F_{\rm tot}$
that come from the bulk-flow kinetic energy flux~$F_{\rm U}$, the AW
enthalpy flux~$F_{\rm w}$, the gravitational potential energy flux~$F_{\rm
  g}$,
the total heat flux~$q_{\rm tot}$, and the enthalpy flux~$F_{\rm e}$.
  \label{fig:fluxratios} }\vspace{0.5cm}
\end{figure}

As discussed in Section~\ref{sec:fluxtube}, the total energy flowing
through the flux tube in our model per unit time, $a F_{\rm tot}$, is
independent of~$r$ in steady state.  (In our numerical solution, the
ratio of the maximum to minimum values of $a F_{\rm tot}$ is~1.003.)
This makes it straightforward to identify the principal source of
energy in our model and to understand how energy is converted from one
form to another as plasma flows away from the Sun.  The total energy
flux defined in Equation~(\ref{eq:Gammatot}) is the sum of the 
bulk-flow kinetic energy flux
\begin{equation}
F_{\rm U} = \frac{1}{2} \rho U^3,
\label{eq:FU} 
\end{equation} 
the gravitational potential energy flux
\begin{equation}
F_{\rm g} = -\,\frac{U G M_{\sun}\rho}{r},
\label{eq:Fg} 
\end{equation} 
the enthalpy flux
\begin{equation}
F_{\rm e} = U nk_{\rm B} \left(\frac{5 T_{\rm e}}{2} + T_{\perp \rm p}
  + \frac{3 T_{\parallel \rm p}}{2} \right),
\label{eq:Fe} 
\end{equation} 
the total heat flux
\begin{equation}
q_{\rm tot} = q_{\rm e} + q_{\perp \rm p} + q_{\parallel \rm p}
\label{eq:qtot} 
\end{equation} 
and the AW enthalpy flux
\begin{equation}
F_{\rm w} = \left(\frac{3U}{2} + v_{\rm A} \right) {\cal E}_{\rm w}.
\label{eq:Fw} 
\end{equation} 
We plot these fluxes in Figure~\ref{fig:fluxratios}, normalized to the
total energy flux~$F_{\rm tot}$. As this figure shows, the wind in
this solution is driven fundamentally by the AW enthalpy flux. As the
plasma flows away from the Sun, part of the AW enthalpy flux is
converted into gravitational potential energy flux as the flow lifts
material out of the Sun's gravitational potential well. Most of the
remaining AW enthalpy flux is gradually converted into bulk-flow kinetic
energy flux, which dominates the total energy flux at $r=1$~AU.

\section{Discussion}
\label{sec:discussion} 

In this section, we discuss our results and
present a second steady-state solution that incorporates
pitch-angle scattering by the cyclotron instability.

\subsection{Solar Rotation}
\label{sec:rotation} 

If solar rotation were taken into account, $\bm{B}$ would follow the
Parker spiral, and at large distances from the axis of rotation
$\bm{B}$ would become approximately azimuthal rather than radial, with
$B\propto r^{-1}$ instead of~$B\propto r^{-2}$. Assuming $n \propto
r^{-2}$, double adiabatic expansion in this azimuthal-field regime
would lead to the scalings $T_{\perp \rm p}\propto r^{-1}$,
$T_{\parallel \rm p}\propto r^{-2}$, and $T_{\perp \rm p}/T_{\parallel
  \rm p} \propto r$.  In contrast, in the radial magnetic field of our
model, double adiabatic expansion leads to $T_{\perp \rm p}\propto
r^{-2}$, $T_{\parallel \rm p} \propto r^0$, and $T_{\perp \rm
  p}/T_{\parallel \rm p} \propto r^{-2}$ assuming $n\propto
r^{-2}$. The inclusion of rotation would thus increase the temperature
anisotropy ratio $T_{\perp \rm p}/T_{\parallel \rm p}$ at large~$r$.
In the solar wind, the transition between radial and azimuthal
magnetic field occurs gradually throughout a range of radii centered
at $r_{\perp} \sim U/\Omega$, where $\Omega$ is the angular frequency of the
Sun's rotation, and $r_\perp$ is distance from the Sun's spin axis.
For fast wind with $U=800$~km/s, and for $\Omega = 2.64 \times 10^{-6}
\mbox{ s}^{-1}$ (the value of~$\Omega$ at a solar latitude
of~$45^\circ$ \citep{snodgrass90}), $U/\Omega = 1.4$~AU.

In addition to modifying the temperature anisotropy ratio at
large~$r$, the inclusion  of rotation would increase the total magnetic field strength
at large~$r$, thereby reducing~$\beta_{\parallel \rm p}$. For example,
during the first polar orbit of Ulysses, the ratio of the mean total
field strength (scaled to $r= \mbox{1 AU}$) to the mean radial field
strength (scaled to $r= \mbox{1 AU}$) was $\simeq
1.7$~\citep{mccomas00}. At fixed $n$ and $T_{\parallel \rm p}$,
increasing $B$ by a factor of $\simeq 1.7$ reduces $\beta_{\parallel
  \rm p}$ by a factor of $\simeq 3$.  Our overestimate
of~$\beta_{\parallel \rm p}$ at $r\sim \mbox{1 AU}$ causes us to
overestimate~$Q_{\parallel \rm p}$ and also pushes the thresholds of
the firehose and mirror instabilities towards smaller temperature anisotropies.

\subsection{The Need for Parallel Proton Cooling}
\label{sec:pparcool} 

In the numerical solution presented in Section~\ref{sec:solution},
$T_{\parallel \rm p} > T_{\perp \rm p}$ at $r > 35 R_{\sun}$. In
contrast, in Helios measurements of fast-wind streams with
$U>700$~km/s, $T_{\perp \rm p}$ typically exceeds $T_{\parallel \rm
  p}$ between $r=60 R_{\sun}$ and $r= 130 R_{\sun}$ \citep{marsch82b}.
Part of this discrepancy may be due to our neglect of solar rotation
(Section~\ref{sec:rotation}) or one or more perpendicular proton
heating mechanisms (Section~\ref{sec:pperpheating}). This discrepancy
may also arise, at least in part, from our neglect of kinetic mechanisms
that act to reduce~$T_{\parallel \rm p}$. For example, part of the
parallel proton thermal energy in the fast solar wind is in the form
of a proton beam~\citep{marsch82b,marsch04}. When the relative speed
of the beam component with respect to the core of the proton
distribution exceeds~$\sim v_{\rm A}$, the proton beam excites
plasma instabilities that slow the beam
down~\citep{daughton98}.  Since $v_{\rm A}$ decreases with
increasing~$r$, these instabilities lead to the steady deceleration of
the proton beam component, which reduces the parallel proton thermal
energy~\citep{hellinger11}. The possible need for parallel proton
cooling in the solar wind was previously suggested by \cite{hu97}.

\subsection{Uncertainties in the Total Turbulent Heating Rate}
\label{sec:turbulence} 

Two of the assumptions in our estimate of~$Q$ become increasingly
inaccurate as $r$ increases. First, we have assumed that $z^-\ll z^+$,
i.e., that most of the AWs propagate away from the Sun in the
solar-wind frame. While $z^+_{\rm rms}$ is likely $\gg z^-_{\rm rms}$
in coronal holes~\citep{cranmer05,verdini07,cranmer10}, Helios
measurements in the fast solar wind show that~$z^+_{\rm rms}/z^-_{\rm
  rms}$ decreases from~$\simeq 4$ to $\simeq 2$ as $r$ increases from
$80 R_{\sun}$ to 1~AU \citep{bavassano00}. Second, the model of
\cite{chandran09c} that we employ to estimate~$z^-_{\rm rms}$ assumes
that the $z^-$ energy cascade time at the outer scale, $t_{\rm casc}^-
\simeq L_\perp/z^+_{\rm rms}$, is much shorter than the linear wave
period~$P$. For the numerical solution presented in
Section~\ref{sec:solution}, $t_{\rm casc}^-$ grows steadily as~$r$
increases, reaching a value $\simeq 3 \times 10^3 \mbox{ s}$ at $r=
100 R_{\sun}$ and a value $\simeq 10^4 \mbox{ s}$ at $r= 200
R_{\sun}$. Thus, for $P \sim 1$~hour, the assumption that $t_{\rm
  casc}^- \ll P$ breaks down at large~$r$.  We also note that velocity
shear may be an important additional source of AW turbulence in the
solar wind, one that is not included in our
model~\citep{roberts87}. As a source of both $z^+$ and $z^-$
fluctuations, AW excitation by velocity shear acts to decrease the
ratio~$z^+_{\rm rms}/z^-_{\rm rms}$ in the solar
wind~\citep{roberts92}.

\subsection{Uncertainties in the Division of the Turbulent Heating Power Between $Q_{\rm
    e}$, $Q_{\perp \rm p}$, and $Q_{\parallel \rm p}$.}
\label{sec:division} 

Our prescription for partitioning the turbulent heating power between
protons and electrons, and between parallel and perpendicular proton
heating, depends strongly upon two of the assumptions we have made in
modeling AW turbulence in the solar wind: the scaling of $\delta
v_{\rm p}$ in Equation~(\ref{eq:dvp}) and the ``critical balance''
condition in Equation~(\ref{eq:critbal}). Equation~(\ref{eq:dvp})
corresponds to an assumption that the inertial-range velocity power
spectrum is $\propto k_\perp^{-3/2}$, which is consistent with
spacecraft measurements at $r = 1 \mbox{
  AU}$~\citep{podesta07,podesta10,chen11} and direct numerical
simulations of AW turbulence in which the fluctuating magnetic field
is $\lesssim 0.2$ times the background magnetic
field~\citep{maron01,muller05,mason06,perez08a}, including simulations
that account for cross helicity, in which $z^+_{\rm rms} > z^-_{\rm
  rms}$ \citep{perez09a}. The power spectrum of the magnetic field is
typically steeper~($\sim k^{-5/3}$) than the velocity power spectrum
in spacecraft
measurements~\citep{matthaeus82,goldstein95a,bruno05,podesta07,chen11}
and direct numerical
simulations~\citep{muller05,boldyrev11a}. The stochastic
heating rate in our model, however, depends upon the velocity power spectrum,
not the magnetic power spectrum, because the velocity spectrum is a
measure of the electric-field fluctuations, which control the
stochastic heating rate when~$\beta_{\parallel \rm p}\lesssim
1$~\citep{chandran10a}. On the other hand, AW turbulence near the Sun
is strongly affected by non-WKB wave reflection and may differ
from AW turbulence at $r\sim \mbox{ 1 AU}$ and from the homogeneous
turbulence in the numerical simulations mentioned above~\citep{verdini09a}. If the
velocity spectrum is steeper (flatter) than we have assumed, then the
amplitude of the velocity fluctuations at $k_\perp \rho_{\rm p} \sim
1$ is smaller (larger) than in Equation~(\ref{eq:dvp}), and stochastic
heating is weaker (stronger) than in our model.

The critical-balance condition that we have adopted in
Equation~(\ref{eq:critbal}) is the same as the condition in Boldyrev's
(2006) theory of scale-dependent dynamic alignment and is consistent
with Perez \& Boldyrev's (2009) \nocite{perez09a} extension of this
theory to the cross-helical case, which holds that the cascade times
of $z^+$ and $z^-$ fluctuations are the same even when $z^+_{\rm rms}
\neq z^-_{\rm rms}$. Equation~(\ref{eq:critbal})  is also consistent
with the work of \cite{podesta10} if the ratio $p/q$ in their analysis
is set equal to unity. However, Equation~(\ref{eq:critbal}) is not
consistent with three competing (and mutually inconsistent) models
of AW turbulence with cross
helicity~\citep{lithwick07,beresnyak08,chandran08a}.
Moreover, none of the six studies just mentioned accounted for the non-WKB
reflection of AWs, which could affect wavenumber anisotropy in the
solar wind. The range of parallel wavenumbers that are present
at~$k_\perp \rho_{\rm p} = 1$ in AW turbulence in the solar wind is
thus uncertain. If the typical values of $|k_\parallel|$ are larger
than we have assumed, then linear wave damping is stronger than we
have assumed, implying that $Q_{\rm \perp p}/Q$ is smaller than in our
model. In addition, near $r=1$~AU, increasing the linear damping rates
would increase $Q_{\parallel \rm p}/Q_{\rm e}$, because less power
would cascade to scales~$\ll \rho_{\rm p}$, and because protons absorb
most of the power that is dissipated at $k_\perp \rho_{\rm p} \sim 1$
when $\beta_{\rm p} > 1$.

Other sources of uncertainty in our partitioning of the turbulent
heating power include the stochastic-heating constant~$c_2$ in
Equation~(\ref{eq:gammas}), which is not known for the case of protons
interacting with strong AW/KAW turbulence.  If we have
underestimated~$c_2$, then we have overestimated the efficiency of
stochastic heating, causing our model to overestimate~$T_{\perp \rm
  p}$. In addition, the use of linear Vlasov theory to estimate
damping rates may be inaccurate when applied to large-amplitude AW/KAW
turbulence in the solar wind~(Borovsky \& Gary~2011; but see also Lehe
et al.~2009). \nocite{borovsky11,lehe09}

\subsection{The Coronal Electron Temperature}
\label{sec:coronalTe} 

Electron temperatures in the low corona inferred from line ratios are
found to be in the range~$\sim 8 \times 10^5 -
10^6$~K~\citep{habbal93,doschek01,wilhelm06,landi08}, similar to the
electron temperature in our model. On the other hand, Ulysses
measurements of ion charge states in the fast solar wind emanating
from the south polar coronal hole suggest that~$T_{\rm e}$ reaches a
maximum of $\simeq 1.5 \times 10^6$~K at $r \sim
1.3-1.5R_{\sun}$~\citep{ko97}. The numerical solution we have
presented in Section~\ref{sec:solution} does not reach such high
electron temperatures, which may indicate that we have under-estimated
the electron heating rate. On the other hand, the electron
temperatures inferred from ion charge-state ratios at
1~AU may be inflated to some degree by the presence of superthermal
electrons in the corona, which could significantly enhance the rates
of ionization into high-energy charge
states~\citep{owocki83,burgi87,ko96}, potentially making the $T_{\rm
  e}$ profile in the left panel of Figure~\ref{fig:swT} consistent
with the ion-charge-state measurements.

\subsection{Other Possible Heating Mechanisms}
\label{sec:pperpheating} 

Although low-frequency AW turbulence is the only non-conductive
heating mechanism in our model, other mechanisms may be important in
the solar wind. For example, compressive
waves are believed to play an important role in chromospheric heating, and
may deposit a significant amount of energy in the low corona as well~\citep{cranmer07,verdini10}.
Type-II spicules may also be an important source of heating in the low corona~\citep{depontieu11}.
Farther from the Sun, the solar wind may be heated by high-frequency
waves that are generated by either a turbulent
cascade~\citep{leamon98a,hollweg02,hamilton08,isenberg11} or by
instabilities driven by the differential flow between the core of the
proton velocity distribution and either alpha particles or proton
beams \citep{gary00,hellinger11}.

\subsection{Cyclotron Instability versus Mirror Instability}
\label{sec:swani_cyclotron} 

\begin{figure*}[t]
\centerline{
\includegraphics[width=5.9cm]{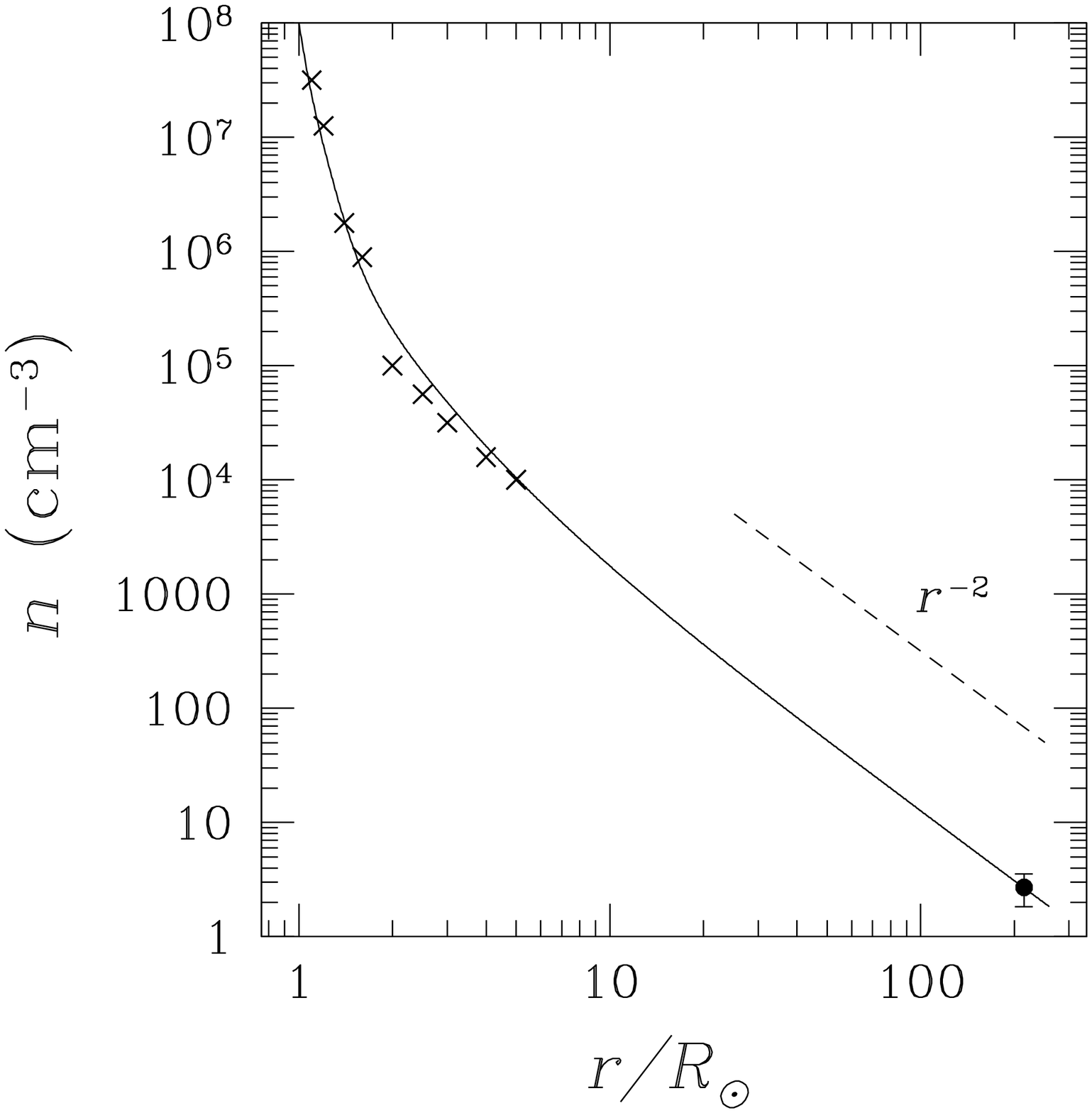}
\hspace{0.2cm} 
\includegraphics[width=5.9cm]{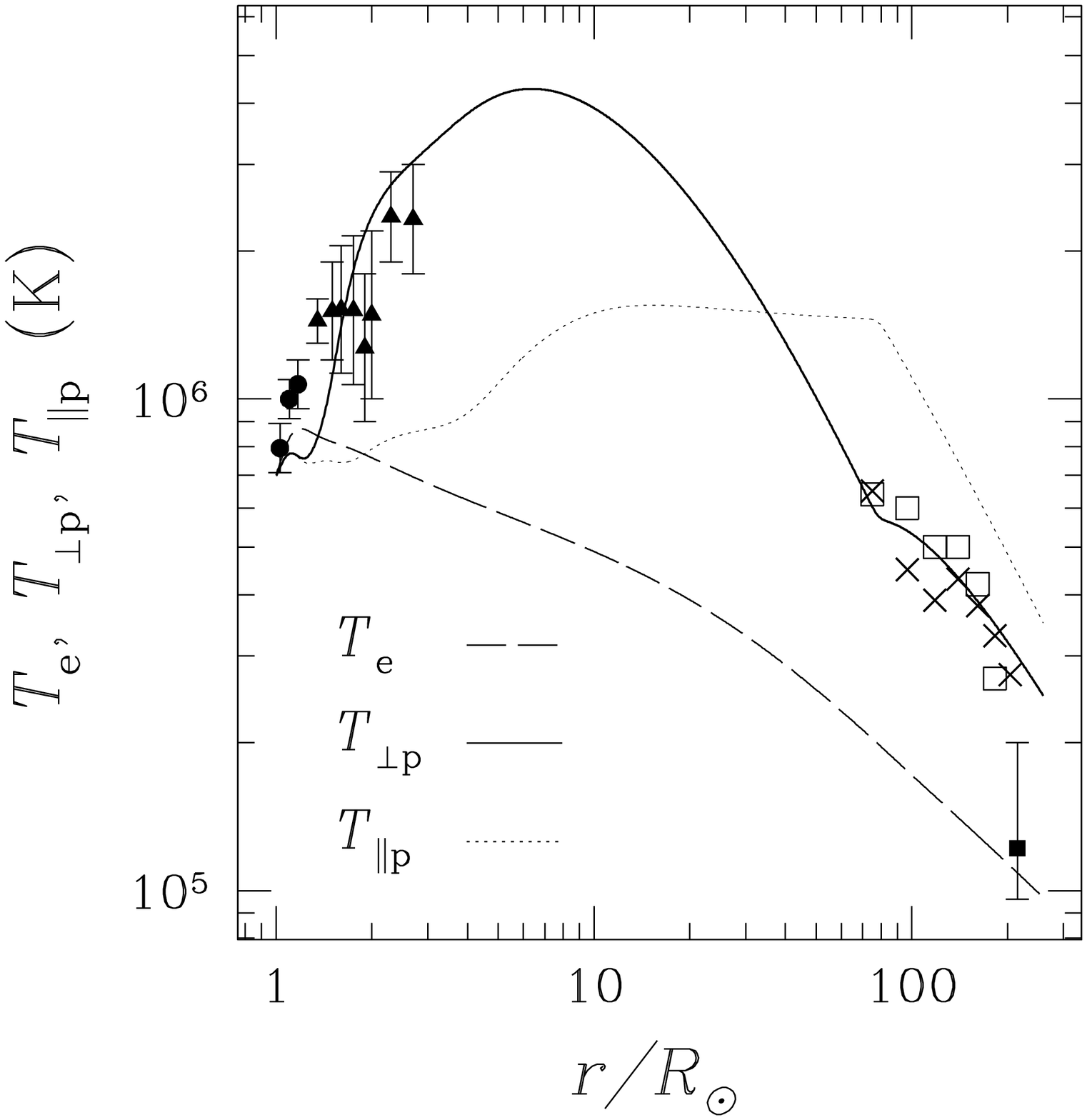}
\hspace{0.2cm} 
\includegraphics[width=5.9cm]{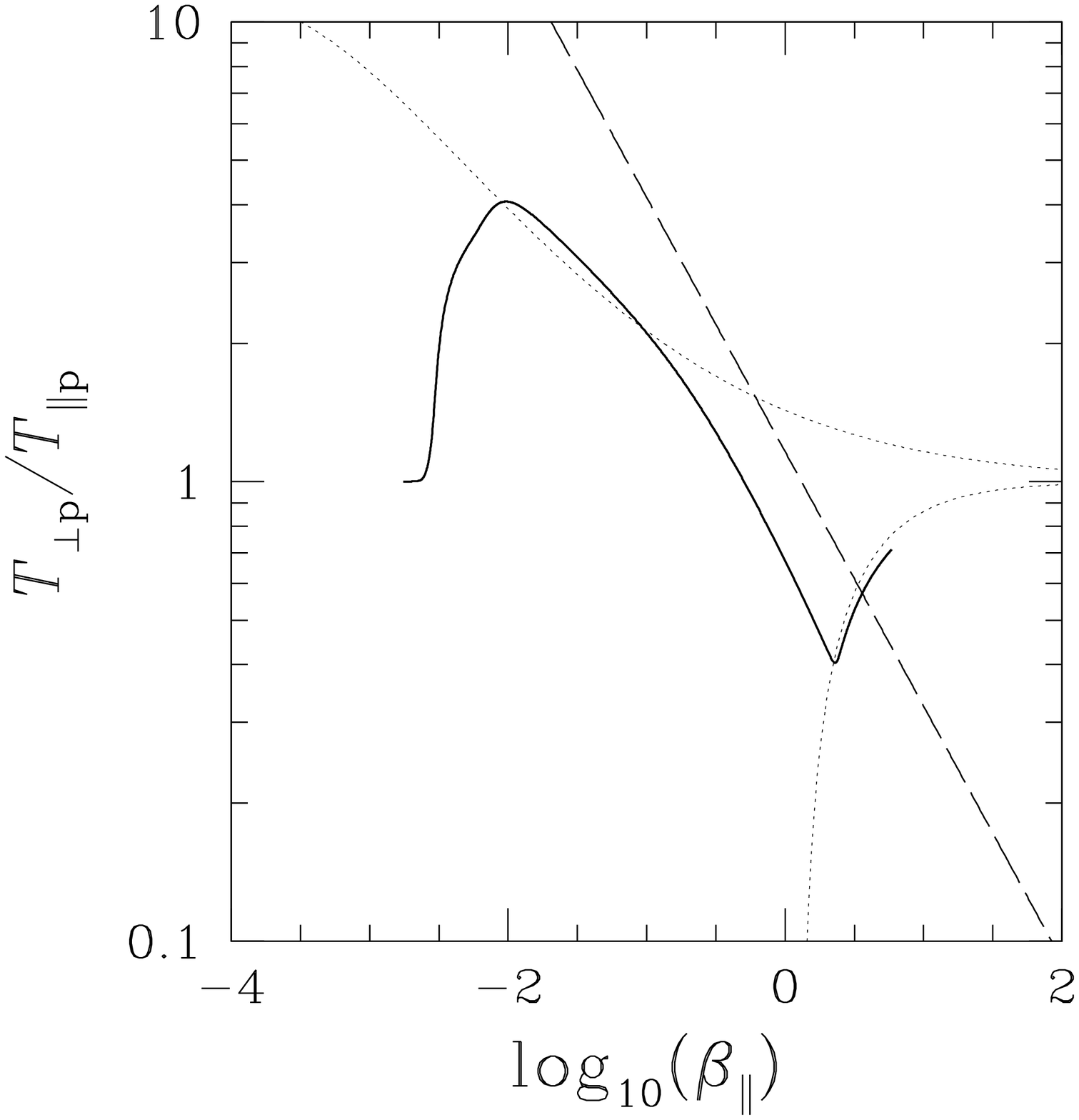}
}
\caption{Steady-state numerical solution when $T_{\perp \rm
    p}/T_{\parallel \rm p}$ is limited from above by the threshold of
  the cyclotron instability (Equation~(\ref{eq:Rcyclotron})) rather
  than the mirror instability. The
  parameter values 
for this solution are given in 
Table~\ref{tab:parameters}.
{\em Left panel:} Same as left panel of Figure~\ref{fig:swn}.
{\em Middle panel:}  Symbols have same meaning as in the middle panel
of Figure~\ref{fig:swT}. In addition, the circles are electron temperatures
inferred from spectroscopic
  observations of a polar coronal hole~\citep{landi08}, and the filled
  square is
  the mean electron temperature in ISEE 3 and Ulysses measurements of
  fast-wind streams with $600 \mbox{ km/s} < U < 700 \mbox{ km/s}$ \citep{newbury98}.
{\em Right panel:}  Same as right panel of Figure~\ref{fig:swT},
except that the upper dotted-line curve is the instability threshold for
  the cyclotron instability in Equation~(\ref{eq:Rcyclotron}).
  \label{fig:cyclotron} }
\vspace{0.5cm} 
\end{figure*}

\cite{hellinger06} showed that near $r=1~AU$, the proton temperature
anisotropy ratio $T_{\perp \rm p}/T_{\parallel \rm p}$ is limited from
above by the threshold of the mirror instability. However, using
Helios measurements, \cite{bourouaine10} found that $T_{\perp \rm
  p}/T_{\parallel \rm p}$ appears to be limited from above by the
cyclotron instability in fast solar wind streams between 0.3~AU and
0.4~AU. To investigate how this latter possibility would affect our
numerical solutions, we have repeated the numerical calculation
presented in Section~\ref{sec:solution} with the parameter values
listed in Table~\ref{tab:parameters} using a model for enhanced proton
pitch-angle scattering based on the cyclotron instability instead of
the mirror instability. That is, we have set $R_{\rm m} \rightarrow
R_{\rm c}$ in Equation~(\ref{eq:nuinst}), with \citep{hellinger06}
\begin{equation}
 R_{\rm c} = 1 + 0.43(\beta_{\parallel \rm p} + 0.0004)^{-0.42} .
\label{eq:Rcyclotron} 
\end{equation} 
The density and temperature profiles in this second steady-state
solution are shown in Figure~\ref{fig:cyclotron}.  The density profile
is very similar to the profile in the left panel of
Figure~\ref{fig:swn}. The profiles of $T_{\rm e}$ and $T_{\perp \rm
  p}$ are very similar to the profiles shown in the left and middle
panels of Figure~\ref{fig:swT}. The most notable difference between
the temperature profiles in the two numerical solutions is
in~$T_{\parallel \rm p}$, which is larger at $4 R_{\sun } \lesssim r
\lesssim 10 R_{\sun}$ when the cyclotron instability threshold is
used. As can be seen in the right panel of Figure~\ref{fig:cyclotron},
when $T_{\perp \rm p}/T_{\parallel \rm p}$ is limited from above by
the anisotropy ratio in Equation~(\ref{eq:Rcyclotron}), the plasma
encounters and then evolves approximately along the
cyclotron-instability threshold at small $r$ and small
$\beta_{\parallel \rm p}$. At larger~$r$ and larger~$\beta_{\parallel
  \rm p}$, the plasma evolves approximately along the threshold of the
oblique firehose instability, as in the numerical solution presented
in Section~\ref{sec:solution}.

\subsection{Sharp Boundaries in Parameter Space}
\label{sec:sharp} 

We have found that there are regions in parameter space in which tiny
changes in certain parameters lead to large changes in the final
steady-state solutions. This phenomenon may be related to the abrupt
transition from fast wind to slow wind in theoretical models in which
the magnetic geometry varies with heliographic
latitude~\citep{cranmer05b,cranmer07}. 

\subsection{Comparison to Previous Studies}
\label{sec:comp}

A number of authors have developed solar-wind models incorporating
temperature
anisotropy~\citep{leer72,whang72,demars91,hu97,olsen99,endeve01,liesvendsen01,janse06},
and some of these models also included energy and/or momentum
deposition by AWs~\citep{hu97,olsen99,liesvendsen01}.  Our work,
however, has several features that are not present in these previous
studies. First, we evaluate the total turbulent heating rate~$Q$ using
an analytical theory of low-frequency AW turbulence driven by non-WKB
wave reflection~\citep{dmitruk02,chandran09c}. Second, we divide the
total turbulent heating rate~$Q$ into three parts ($Q_{\rm e}$,
$Q_{\perp \rm p}$ and $Q_{\parallel \rm p}$) using an analytical model
of the collisionless dissipation of low-frequency AW
turbulence. Third, we do not include any heating other than that which
is provided by the proton and electron heat fluxes and low-frequency
AW turbulence. Fourth, we account for the mirror (or cyclotron) and
oblique firehose instabilities by enhancing the proton pitch-angle
scattering rate when the proton temperature anisotropy exceeds the
threshold of either instability.

Two other groups of authors have developed solar-wind models based on
energy and momentum deposition by low-frequency AW turbulence driven
by non-WKB wave reflection~\citep{cranmer07,verdini10}. These authors
allowed for a broad spectrum of AW frequencies at the coronal base and
accounted for the fact that higher-frequency waves undergo less
reflection. In contrast, our estimates for $z^-_{\rm rms}$ and~$Q$ in
Equations~(\ref{eq:zminus}) and (\ref{eq:Q}) were obtained in the
low-frequency limit~\citep{chandran09c}. On the other hand, these
models treated the solar wind as a single fluid, without
distinguishing between proton and electron temperatures. Also, neither
model incorporated temperature anisotropy.

\vspace{0.2cm} 
\section{Conclusion}
\label{sec:conclusion} 
\vspace{0.2cm} 

We have developed a two-fluid model of the solar wind that accounts
for proton temperature anisotropy, pitch-angle scattering from mirror
(or cyclotron) and firehose instabilities, and (kinetic) Alfv\'en Wave
(AW/KAW) turbulence. We neglect rotation and consider solar wind
flowing along a narrow magnetic flux tube centered on a radial
magnetic field line. The turbulent heating in our model is partitioned
between electrons and protons, and between perpendicular and parallel
proton heating, in accord with recent results on stochastic ion
heating~\citep{chandran10a} and linear damping rates calculated from
the full hot-plasma dispersion relation.  The electron heat flux in
our model transitions from the Spitzer-H\"arm value in the collisional
region near the Sun to the \cite{hollweg76} value in the nearly
collisionless conditions at larger~$r$. To evaluate the proton heat
flux in the presence of temperature anisotropy in both collisional and
collisionless conditions, we use a fourth-velocity-moment fluid
closure based on the guiding-center Vlasov equation
\citep{kulsrud83,snyder97}. Our model conserves energy, and in steady
state the total energy flowing through the cross section of the flux
tube becomes independent of~$r$. No energy is added to the solution
through ad hoc heating terms.

In Section~\ref{sec:solution} we present a steady-state numerical
solution to our model equations for $R_{\sun} < r < 1.2 \mbox{ AU}$,
which we analyze in detail in order to gain insight into the different
physical processes operating within the model. As shown in
Figure~\ref{fig:fluxratios}, it is the AW enthalpy flux~$F_{\rm w}$
that drives the solar wind in this solution. As plasma flows away from
the Sun, the AW enthalpy flux is gradually converted into
gravitational potential energy flux and bulk-flow kinetic energy
flux. By the time the flow reaches 1~AU, the total energy flux is
dominated by the bulk-flow kinetic energy.  As shown in
Figure~\ref{fig:swQe}, electrons are heated primarily by the
dissipation of AW/KAW turbulence at $r\leq 2.5 R_{\sun}$.  At
$r\gtrsim 3 R_{\sun}$, AW turbulence and conduction provide comparable
amounts of electron heating, to within a factor of~$\sim 2$.  Between
$1.35 R_{\sun}$ and $185 R_{\sun}$, AW/KAW turbulence dissipates
primarily via stochastic proton heating, leading to substantial
perpendicular proton heating, as shown in the right panel of
Figure~\ref{fig:swdv}. Parallel proton heating via Landau and
transit-time damping accounts for $< 10\%$ of the total turbulent
heating at all radii in this solution, and $Q_{\parallel \rm p} < 0.01
Q$ at $r< 100 R_{\sun}$. The proton heat fluxes $q_{\perp \rm p}$ and
$q_{\parallel \rm p}$ cause $T_{\parallel \rm p}$ to increase
gradually as~$r$ increases from~$2R_{\sun}$ to $\sim 20 R_{\sun}$,
despite the fact that solar-wind expansion reduces $q_{\perp \rm p}$
and $q_{\parallel \rm p}$ relative to their free-streaming values.  At
$r= 71 R_{\sun}$, the plasma encounters the threshold of the oblique
firehose instability, causing the proton pitch-angle scattering rate
to increase. At $r> 71 R_{\sun}$, the proton temperature anisotropy
ratio evolves approximately along the firehose-instability threshold
as $\beta_{\parallel \rm p}$ increases.

This numerical solution is broadly consistent with a number of
observations, as illustrated in Figures~\ref{fig:swn}, \ref{fig:swdv},
\ref{fig:swT}, and~\ref{fig:swq}, supporting the idea 
that AW turbulence may be one of the primary
mechanisms responsible for heating coronal holes and accelerating the
solar wind~\citep{parker65,coleman68}.  Perhaps the most notable
achievement of our model is that it comes close to explaining
observations of perpendicular proton temperatures, even though the
heating in the model is provided by low-frequency AW turbulence rather
than resonant cyclotron interactions. The main uncertainties in our
results are associated with the stochastic heating rate in strong
AW/KAW turbulence, the wavenumber anisotropy ($k_\parallel /k_\perp$)
in reflection-driven AW/KAW turbulence, the total turbulent heating rate at
large~$r$, and the effects of solar rotation on the temperature
anisotropy ratio~$T_{\perp \rm p}/T_{\parallel \rm p}$.

One of our principal objectives in this work has been to to connect
theoretical studies of microphysical processes with observations of
macrophysical quantities in the solar wind. By comparing our model to
observations, we have been able to obtain a new test of the viability
of AW turbulence as a mechanism for heating the solar wind and coronal
holes. At this point, the results of this test are encouraging, but
not fully conclusive, because of the uncertainties described above.
However, as our understanding of kinetic plasma physics and turbulence
in the solar wind progresses, it will be possible to use models such
as the one we have developed to obtain increasingly rigorous tests of
competing theories and, ultimately, to gain greater insight into
coronal heating and the origin of the solar wind.

\acknowledgements We thank Steve Cranmer, Joe Hollweg, Greg Howes, Phil Isenberg,
Yuan-Kuen Ko, Jean Perez, and Jason Tenbarge for helpful discussions. This work
was supported in part by grant NNX11AJ37G from NASA's Heliophysics
Theory Program, NSF grant AGS-0851005, NSF/DOE grant AGS-1003451, DOE
grant DE-FG02-07-ER46372, NSF/DOE grant PHY-0812811, and NSF grant
ATM-0752503.

\appendix

\vspace{0.2cm} 
\section{Fourth-Moment Fluid Closure of the Guiding-Center Vlasov Equation}
\label{ap:eqns} 
\vspace{0.2cm} 

The derivation of the equations in our model begins with Kulsrud's formulation
of collisionless MHD for a proton-electron plasma \citep{kulsrud83}.
Kulsrud's approach was to expand all quantities in the Vlasov and
Maxwell equations in powers of $1/e$, where $e$ is the proton charge,
and to consider the limit~$e\rightarrow \infty$.  This limit
corresponds to the case in which the Debye length~$\lambda_{\rm D}$
and proton gyroradius~$\rho_{\rm p}$ are much smaller than the length
scales over which the macroscopic quantities vary appreciably.  The
fundamental variables in Kulsrud's theory are the mass
density~$\rho$, the fluid velocity~$\bm{U}$ (which is the same for
electrons and protons to lowest order in~$1/e$), the magnetic
field~$\bm{B}$, the proton and electron distribution functions~$f_{\rm
  p}$ and $f_{\rm e}$, and the parallel component of the electric
field~$\bm{E}$, given by $E_\parallel = \bm{\hat{b}} \cdot \bm{E}$,
where $\bm{\hat{b}} = \bm{B}/B$.  To lowest order in~$1/e$, these
variables satisfy the equations
\begin{equation}
\frac{\partial \rho}{\partial t} + \bm{\nabla} \cdot (\rho \bm{U}) = 0,
\label{eq:cont0} 
\end{equation} 
\begin{equation}
\rho\left(\frac{\partial \bm{U}}{\partial t} +
  \bm{U} \cdot \bm{\nabla}\bm{U}\right) = \frac{(\bm{\nabla}\times
  \bm{B}) \times \bm{B}}{4\pi} - \bm{\nabla} \cdot \bm{P},
\label{eq:momentum0} 
\end{equation} 
\begin{equation}
\frac{\partial \bm{B}}{\partial t} = \bm{\nabla} \times (\bm{U} \times
\bm{B}),
\label{eq:induction} 
\end{equation} 
\begin{equation}
\bm{P} = \sum_{s}p_{\perp s} (\bm{I} - \bm{\hat{b}}\bm{\hat{b}}) +
\sum_{s}p_{\parallel s} \bm{\hat{b}}\bm{\hat{b}},
\label{eq:defPtensor} 
\end{equation} 
\begin{equation}
p_{\perp s} = \frac{m_s}{2} \int f_s\, |\bm{v} - v_\parallel
\bm{\hat{b}} - \bm{v}_E |^2 d^3 v,
\label{eq:defpperp} 
\end{equation} 
\begin{equation}
p_{\parallel s} = m_s \int f_s\, (v_\parallel - U_\parallel)^2 d^3 v,
\label{eq:defppar} 
\end{equation} 
\begin{equation}
\sum_{s} e_s n_s = 0,
\label{eq:quasineutrality} 
\end{equation} 
\begin{equation}
n_s = \int f_s\, d^3 v ,
\label{eq:defns} 
\end{equation} 
and (see Snyder et al.~1997)
\begin{equation}
\frac{\partial}{\partial t}(f_{\rm s} B) \;+\; \bm{\nabla} \cdot [
f_{\rm s} B(v_\parallel \bm{\hat{b}} + \bm{v}_{E})]
+ \frac{\partial
}{\partial v_{\parallel}}\left[ f_{\rm s} B \left( - \bm{\hat{b}}
    \cdot \frac{D \bm{v}_{E}}{Dt} - \mu \bm{\hat{b}} \cdot
    \bm{\nabla}B + \frac{e_s E_\parallel}{m_s} \right)\right]
\; = \;0,
\label{eq:dfdt} 
\end{equation} 
where $s$ is a subscript indicating particle species (p for proton and
e for electron), $f_s$ is the distribution function of particle
species~$s$, $m_s$ and $e_s$ are the mass and charge of species~$s$,
$\bm{v}$ is particle velocity, $v_\parallel = \bm{\hat{b}} \cdot
\bm{v}$, $\bm{v}_{E} = c (\bm{E}\times\bm{B})/B^2$, $\mu = |\bm{v} -
v_\parallel \bm{\hat{b}} - \bm{v}_{E}|^2/2B$, $U_\parallel = \bm{U}
\cdot \bm{\hat{b}}$, and $D/Dt = \partial /\partial t + (v_\parallel
\bm{\hat{b}} + \bm{v}_{E}) \cdot \bm{\nabla}$. In
Equation~(\ref{eq:dfdt}), $f_s$ is regarded as a function of
position~$\bm{x}$, time~$t$, magnetic moment~$\mu$, and parallel
velocity~$v_\parallel$.  Rather than retain the subscripts on the
number densities, we define
\begin{equation}
n = n_{\rm p},
\label{eq:defn} 
\end{equation} 
which is also equal to~$n_{\rm e}$ because of
Equation~(\ref{eq:quasineutrality}).  We
have neglected the electron contribution to the mass density, setting
$\rho = n m_{\rm p}$.
The parallel and perpendicular temperatures are related to the parallel
and perpendicular pressures defined above in the usual way:
$p_{\perp \rm s} = n k_{\rm B} T_{\perp \rm s}$ and
$p_{\parallel \rm s} = n k_{\rm B} T_{\parallel \rm s}$.

\cite{snyder97} extended Kulsrud's collisionless MHD 
to account for collisions by adding a BGK collision operator
\citep{gross56} to the right-hand side of
Equation~(\ref{eq:dfdt}). For the case we consider here, in which the
electrons and ions have the same average velocity~$\bm{U}$ and number
density~$n$, this collision operator takes the form
\begin{equation}
C(f_s) = \sum_{k} \nu_{sk} (F_{M s} - f_s),
\label{eq:Cf} 
\end{equation} 
where 
\begin{equation}
F_{Ms} = n \left(\frac{m_s}{2 \pi k_{\rm B} T_{\rm s}}\right)^{3/2}
  \exp\left[ - \frac{m_{\rm s}(v_\parallel - U_\parallel)^2}{2 k_{\rm
        B} T_s} -
    \frac{m_s \mu B}{k_{\rm B} T_s}\right],
\label{eq:FM} 
\end{equation} 
is a shifted Maxwellian with temperature
\begin{equation}
T_s = \frac{2 T_{\perp s} + T_{\parallel s}}{3},
\label{eq:Ts} 
\end{equation} 
and $\nu_{sk}$ is the collision frequency for momentum exchange
between species $s$ and species~$k$. (Here, we neglect energy exchange
between protons and electrons, but we include it in
section~\ref{sec:fluxtube}.)  \cite{snyder97} then obtained a
hierarchy of fluid equations by multiplying Equation~(\ref{eq:dfdt})
by various powers of~$v_\parallel$ and~$\mu$ and then integrating over
$v_\parallel$ and~$\mu$. For the protons, the equations for $p_{\perp
  \rm p}$ and $p_{\parallel \rm p}$ can be
written~\citep{snyder97,sharma06a}
\begin{equation}
\rho B\frac{d}{dt}\left(\frac{p_{\perp \rm p}}{\rho B}\right) =  -
\bm{\nabla}\cdot(q_{\perp \rm p}\bm{\hat{b}}) - q_{\perp \rm p}
\bm{\nabla} \cdot\bm{\hat{b}} + \frac{\nu_{\rm p}}{3} (p_{\parallel
  \rm p} - p_{\perp \rm p})
\label{eq:dpperpdt1} 
\end{equation} 
and
\begin{equation}
\frac{\rho^3}{2B^2}\frac{d}{dt}\left(\frac{B^2 p_{\parallel \rm p}
    }{\rho^3}\right) =  - \bm{\nabla}\cdot(q_{\parallel \rm
  p}\bm{\hat{b}}) +  q_{\perp \rm p} \bm{\nabla}\cdot\bm{\hat{b}}
+ \frac{ \nu_{\rm p}}{3} (p_{\perp
  \rm p} - p_{\parallel \rm p}),
\label{eq:dppardt1} 
\end{equation} 
where
\begin{equation}
q_{\perp \rm p} = m_{\rm p} \int f_{\rm p}\, \mu B (v_\parallel - U_\parallel) d^3 v,
\label{eq:defqperpp} 
\end{equation} 
\begin{equation}
q_{\parallel \rm p} = \frac{m_{\rm p}}{2} \int f_{\rm p}\, (v_\parallel - U_\parallel)^3 d^3 v,
\label{eq:defqparp} 
\end{equation} 
$m_{\rm p}$ is the proton mass, 
and
\begin{equation}
\nu_{\rm p} = \nu_{\rm pp} + \nu_{\rm pe}.
\label{eq:nup} 
\end{equation} 
In Section~\ref{sec:fluxtube} we neglect the $\nu_{\rm pe}$ term in
Equation~(\ref{eq:nup}) because it is smaller than the proton-proton Coulomb
collision frequency. Our $q_{\parallel \rm p}$ is by definition a
factor of~2 smaller than Snyder et al.'s (1997).  In this appendix,
the Lagrangian time derivative is given by
\begin{equation}
\frac{d}{dt} = \frac{\partial}{\partial t} + \bm{U} \cdot \bm{\nabla},
\label{eq:lagder0} 
\end{equation} 
which generalizes Equation~(\ref{eq:lagder1}) to allow for arbitrary
flow velocities.
The equations for $q_{\perp \rm p}$ and $q_{\parallel \rm p}$ are
\begin{equation}
\rho^2 \frac{d}{dt} \left(\frac{q_{\perp \rm p}}{\rho^2}\right)  +
\nu_{\rm p} q_{\perp \rm p} = 
- \bm{\nabla} \cdot (r_{\parallel \perp} \bm{\hat{b}}) +
\frac{p_{\perp \rm p}}{\rho}\, \bm{\hat{b}}\cdot \bm{\nabla}p_{\parallel
  \rm p} 
+ \left[ \frac{p_{\perp \rm p}(p_{\parallel \rm p} - p_{\perp
      \rm p})}{\rho} + r_{\perp \perp} - r_{\parallel \perp}\right]
\bm{\nabla} \cdot \bm{\hat{b}}
\label{eq:dqperpdt1} 
\end{equation} 
and
\begin{equation}
\frac{\rho^4}{B^3} \frac{d}{dt}\left(\frac{B^3 q_{\parallel \rm
      p}}{\rho^4}\right)  + \nu_{\rm p} q_{\parallel \rm p} =
- \frac{1}{2} \bm{\nabla}\cdot(r_{\parallel \parallel}\bm{\hat{b}}) +
\frac{3p_{\parallel \rm p}}{2\rho}\,\bm{\hat{b}}\cdot \nabla
p_{\parallel \rm p}
+ \frac{3}{2}\left[\frac{p_{\parallel \rm p}(p_{\parallel \rm p} - p_{\perp \rm
      p})}{\rho} + r_{\parallel \perp}\right]\bm{\nabla}\cdot\bm{\hat{b}},
\label{eq:dqpardt1}
\end{equation} 
where
\begin{equation}
r_{\perp \perp} = m_{\rm p} \int f_{\rm p} \,\mu^2 B^2 d^3 v,
\label{eq:rperpperp} 
\end{equation} 
\begin{equation}
r_{\parallel \perp} = m_{\rm p} \int f_{\rm p}\, \mu B (v_\parallel - U_\parallel)^2 d^3 v,
\label{eq:rparperp} 
\end{equation} 
and 
\begin{equation}
r_{\parallel \parallel} = m_{\rm p} \int f_{\rm p}\, (v_\parallel - U_\parallel)^4 d^3 v,
\label{eq:rparpar} 
\end{equation} 
In Equation~(\ref{eq:dqperpdt1}), we have corrected a minor
error in Equation~(19) of \cite{snyder97}: the fourth term on the
left-hand side of their Equation~(19) should be~$q_{\perp \rm s} \bm{\nabla}
\cdot \bm{U}$ instead of $q_{\perp \rm s}\bm{\nabla} \cdot
(U_\parallel \bm{\hat{b}})$.

To close these fluid equations, we set 
$f_{\rm
  p}= F_{BM}$ in Equations~(\ref{eq:rperpperp}) through (\ref{eq:rparpar}), where 
\begin{equation}
F_{BM} = \frac{n m_{\rm p}^{3/2}}{(2\pi k_{\rm B})^{3/2} T_{\perp
    \rm p} T_{\parallel \rm p}^{1/2}} \exp\left[ - \frac{m_{\rm p}
    \mu B}{ k_{\rm B} T_{\perp p}} - \frac{m_{\rm p}
    (v_\parallel - U_\parallel)^2}{2 k_{\rm B}
    T_{\parallel \rm p}}\right]
\label{eq:FBM} 
\end{equation} 
is a bi-Maxwellian distribution with the same values of~$n$,
$U_\parallel$, $T_{\perp \rm p}$ and $T_{\parallel \rm p}$ as the
protons. This enables us to rewrite Equations~(\ref{eq:dqperpdt1}) and
(\ref{eq:dqpardt1}) as
\begin{equation}
\rho^2 \frac{d}{dt} \left(\frac{q_{\perp \rm p}}{\rho^2}\right) +
\nu_{\rm p} q_{\perp \rm p} = 
- \frac{n k_{\rm B}^2 T_{\parallel \rm p}}{m_{\rm p}}\, \bm{\hat{b}}
\cdot \bm{\nabla} T_{\perp \rm p}
 + \frac{n k_{\rm B}^2 T_{\perp \rm
    p} (T_{\perp \rm p} - T_{\parallel \rm p})}{m_{\rm p}}\,
\bm{\nabla} \cdot
    \bm{\hat{b}}
\label{eq:dqperpdt2} 
\end{equation} 
and
\begin{equation}
\frac{\rho^4}{B^3} \frac{d}{dt}\left(\frac{B^3 q_{\parallel \rm
      p}}{\rho^4}\right) + \nu_{\rm p} q_{\parallel \rm p} =
- \frac{3 n k_{\rm B}^2 T_{\parallel \rm p}}{2m_{\rm p}} \,
\bm{\hat{b}}\cdot \bm{\nabla}T_{\parallel \rm p}.
\label{eq:dqpardt2} 
\end{equation}
Equations~(\ref{eq:dqperpdt2}) and (\ref{eq:dqpardt2}) were derived
in different ways and with differing treatments of collisions
by~\cite{endeve01}, \cite{liesvendsen01}, and~\cite{ramos03}.

For the electrons, we set $T_{\perp \rm e} = T_{\parallel \rm e} =
T_{\rm e}$ as described in section~\ref{sec:fluxtube}, and close the
electron fluid equations by specifying the electron heat
flux~$\bm{q}_{\rm e}$ in terms of lower moments of the electron
distribution (Section~\ref{sec:eheatflux}). The electrons then
satisfy a standard energy equation,
\begin{equation}
\frac{3}{2} n^{5/3} k_{\rm B} \frac{d}{dt}\left(\frac{T_{\rm e}}{n^{2/3}}\right) =
- \, \bm{\nabla}\cdot \bm{q}_{\rm e} .
\label{eq:dTedt0} 
\end{equation} 
In Section~\ref{sec:fluxtube}, we adapt the general equations given in
this appendix to our 1D solar-wind model and add extra terms to
incorporate the effects of AW turbulence, collisional
energy exchange between protons and electrons, and
temperature isotropization by firehose and mirror instabilities.

\bibliography{articles}

\end{document}